\documentclass[letterpaper,twocolumn,10pt]{article}
\PassOptionsToPackage{dvipsnames}{xcolor}
\usepackage{usenix2019_v3}
\usepackage{endnotes}
\usepackage{graphicx}
\usepackage{tabularx}
\usepackage{multirow, booktabs}

\usepackage{enumerate}
\usepackage{enumitem}
\usepackage{subcaption}
\usepackage{amsmath}
\usepackage[ruled,linesnumbered,vlined]{algorithm2e}
\usepackage{listings}

\usepackage{amssymb}
\usepackage{pifont}
\usepackage{xspace}

\definecolor{dkgreen}{rgb}{0,0.6,0}
\definecolor{gray}{rgb}{0.5,0.5,0.5}
\definecolor{mauve}{rgb}{0.58,0,0.82}
\definecolor{codegreen}{rgb}{0,0.6,0}
\definecolor{codegray}{rgb}{0.5,0.5,0.5}
\definecolor{codepurple}{rgb}{0.58,0,0.82}
\definecolor{backcolour}{rgb}{0.95,0.95,0.92}
\definecolor{ACMBlue}{HTML}{0072CE}

\usepackage{makecell}

\usepackage{framed}

\usepackage{cleveref}\crefname{section}{\S}{\SS}
\crefformat{section}{\S#2#1#3} \crefformat{subsection}{\S#2#1#3}
\crefformat{subsubsection}{\S#2#1#3}

\setlist{itemsep=0pt,parsep=0pt,topsep=0pt}

\usepackage[compact]{titlesec}
\usepackage[font=small,labelfont=bf,aboveskip=8pt]{caption}
\usepackage[all]{nowidow}

\usepackage{tikz}
\usetikzlibrary{backgrounds,positioning,fit,calc}

\usepackage[most]{tcolorbox}
\definecolor{light-gray}{gray}{0.95}
\newtcolorbox{observation}{
    center,
    width=\linewidth,
    colframe=light-gray,
    colback=light-gray
}

\usepackage{wrapfig}
\usepackage{xurl}

\definecolor{americanrose}{rgb}{1.0, 0.01, 0.24}
\definecolor{darkgreen}{rgb}{0.0, 0.5, 0.0}

\DeclareRobustCommand*\circled[1]{\tikz[baseline=(char.base)]{\node[shape=circle,fill,minimum size=1.07em,inner sep=0pt] (char) {\footnotesize\textcolor{white}{#1}};}}

\newcommand{\system}{\textsf{LatentStore}\xspace}
\newcommand{\civitai}{\textsf{Civitai}\xspace}

\newcommand{\rev}[1]{\textcolor{black}{#1}}

\newcommand{\phead}[1]{\noindent\textbf{#1}} \newcommand{\fref}[1]{Fig.~\ref{#1}}
\newcommand{\tref}[1]{Table~\ref{#1}}
\newcommand{\eref}[1]{Eq.~\ref{#1}}
\newcommand{\sref}[1]{\cref{#1}}

\title{{\system}: An Efficient Latent-First Storage System for AI-Generated Images} 

\author{
Zirui Wang$^1$, Yunjia Zheng$^2$, Tingfeng Lan$^1$, Zhaoyuan Su$^1$, Haoran Ni$^2$, Juncheng Yang$^2$, Yue Cheng$^1$ \\
$^1$University of Virginia, $^2$Harvard University
}
\begin{document}
\maketitle
\begin{abstract}

The explosive growth of AI-generated images has created a sustainability challenge for storage infrastructure. 
Platforms like Midjourney and Adobe Firefly already host billions of generative images, yet conventional object stores persist them as blobs with full-resolution pixels, consuming huge amounts of storage capacity and bandwidth. 
Unlike natural photos, however, 
AI-generated images can be deterministically reconstructed from compact, model-native latent tensors, 
making persistent image storage fundamentally redundant.  

This paper presents {\system}, a latent-first storage system for AI-generated images. 
{\system} treats compressed latents as durable storage objects and uses 
on-demand GPU reconstruction on the read path to trade inexpensive compute for large persistent storage savings.  
Our design is guided by the first large-scale analysis of AI-generated image access we are aware of, based on a 35-month, 2-billion-request production trace from a major generative-content platform. Motivated by the trace analysis, 
{\system} keeps frequently accessed images in decoded pixel format for fast hits, stores less-active objects as compressed latents to expand effective cache capacity, and continuously adjusts the splits between the image and latent cache to optimize user-perceived access latency. 
We build a {\system} prototype and evaluate it with the production trace. 
{\system} reduces persistent storage by 78.7\% with competitive or even lower mean and tail latency over a pure image-based storage.

\end{abstract}
\section{Introduction}
\label{sec:introduction}

The proliferation of text-to-image generative models has created an unprecedented volume of synthetic visual content.
Platforms such as Midjourney~\cite{midjourney2026} now produce over 34~million images per day~\cite{everypixel2023aiimagestats}, while Adobe Firefly has generated more than 29~billion images since its 2023 launch~\cite{adobe2023fireflyfoundry}.
At typical resolutions ($1024{\times}1024$), storing 100~billion losslessly compressed PNGs requires roughly 150\,PB, with costs scaling linearly as new images are created every day, making storage a dominant and fast-growing infrastructure expense for generative-content platforms.

\begin{figure}[t]
\centering
\begin{subfigure}[t]{0.585\columnwidth}
\includegraphics[width=\linewidth]{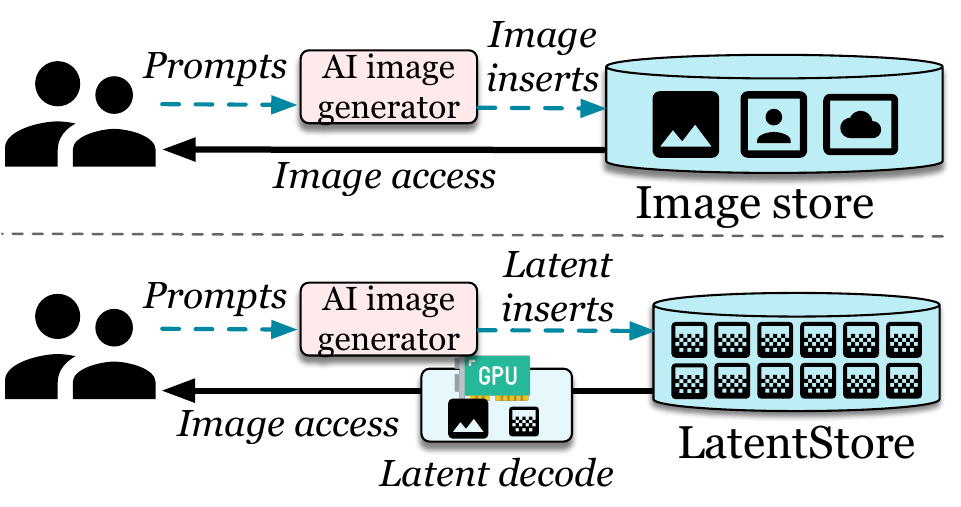}
\vspace{-15pt} 
\caption{\rev{Conventional image store vs. {\system}.}}\label{fig:ls_conceptual} 
\end{subfigure}
\hfill
\begin{subfigure}[t]{0.395\columnwidth} 
\includegraphics[width=\linewidth]{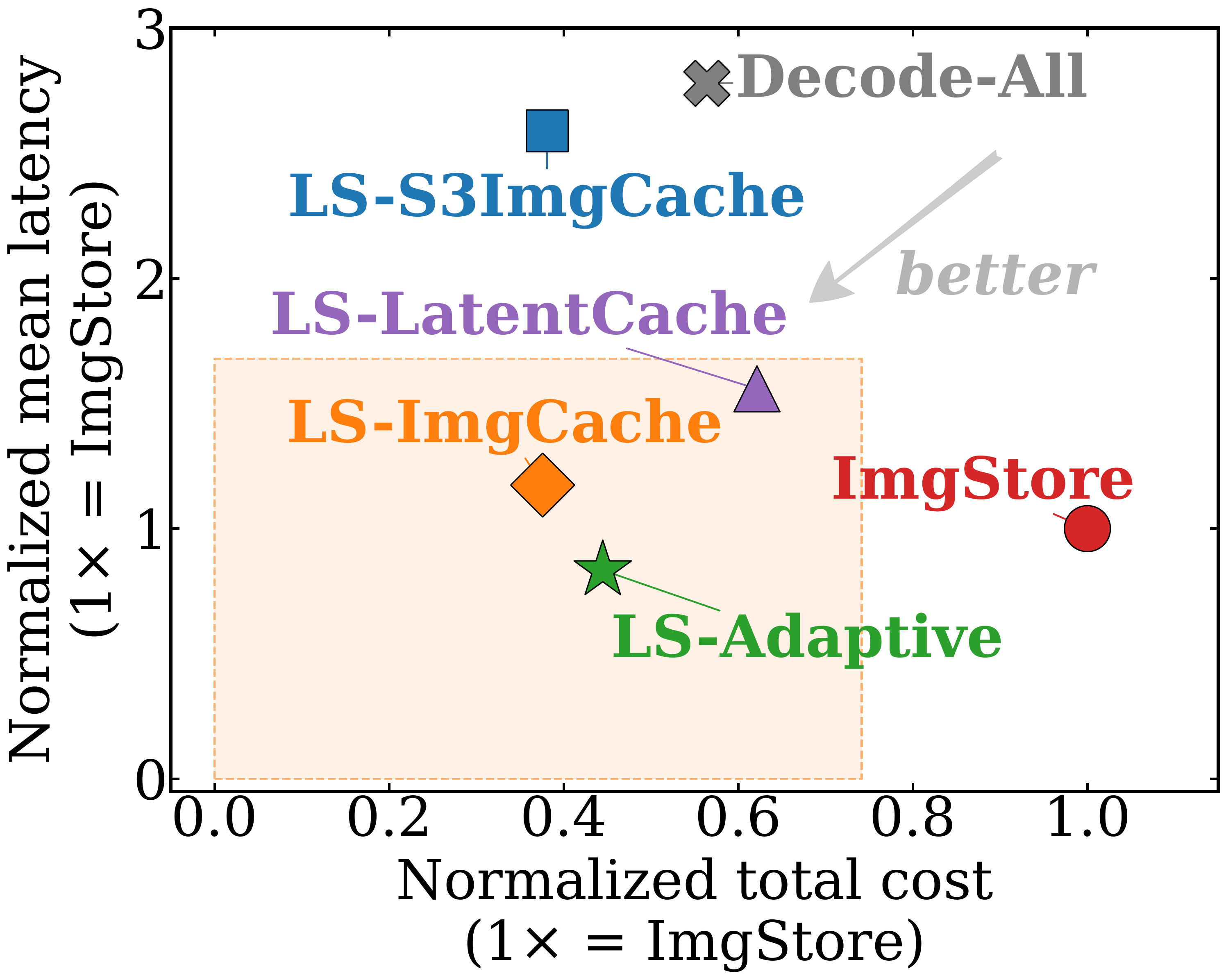}
\vspace{-15pt}
\caption{\rev{Cost--latency tradeoff.}}\label{fig:intro_scatter}
\end{subfigure}
\vspace{-5pt}
\caption{Illustration of latent-first storage.
(a) Conventional object stores persist AI-generated images as opaque blobs, whereas \system (LS) stores compact model-native latents (intermediate state) and reconstructs images on demand.
(b) Cost--latency tradeoff of \rev{six} storage strategies. {\system} achieves low cost and latency. 
}
\label{fig:ls_concept_scatter}
\vspace{-5pt}
\end{figure}

Existing large-scale image storage systems, including Facebook's Haystack~\cite{beaver2010finding} and f4~\cite{muralidhar2014f4}, LinkedIn's Ambry~\cite{noghabi2016ambry}, and Tencent's photo store~\cite{tencent_trace}, treat images as opaque blobs.
Their optimization levers are limited to \emph{where} to place blobs, \emph{how} to replicate them, and \emph{what} to cache.
These systems cannot exploit a property unique to AI-generated content: every image is a 
\emph{deterministic function of a compact, model-native latent tensor~\cite{rombach2022high}}, \rev{i.e., the low-dimensional representation that a diffusion model produces by iterative denoising and that its VAE decoder maps to pixels}, making persistent storage of full-resolution images redundant.

Modern diffusion models produce images through a two-stage pipeline: an expensive iterative diffusion process that takes seconds to generate a small latent tensor, followed by a lightweight VAE decoder~\cite{kingma2013auto,rezende2014stochastic,esser2021taming,van2017neural,chen2024deep} that maps the latent to images in tens of milliseconds. 
The decode is deterministic (the same latent always yields the same image) and the latent is roughly $5{\times}$ smaller than the corresponding PNG. 
This asymmetry reveals the key insight of this paper: 
\emph{Image generation is expensive but performed only once.
Storing the compact latent minimizes persistent storage cost, while image reconstruction via GPU decode is cheap enough to be performed on demand without inflating user-perceived latency.} 

A \emph{latent-first} storage system can therefore replace stored images with much smaller latents and reconstruct on the fly.
However, realizing the storage savings without sacrificing user-facing performance introduces new challenges: requests that cannot be served from a decoded image cache now triggers a GPU decode rather than a simple storage read.  

Analysis of a large production trace from a major generative-content platform {\civitai}~\cite{civitai} shows that image popularity is heavily skewed yet rapidly decaying, re-access intervals span seconds to months, and even a well-sized cache leaves a persistent miss residual. 
These properties make a latent-first design nontrivial. 
If the system caches decoded images, popular requests are fast because they avoid GPU decode, but the cache holds fewer objects and more requests fall through to slower paths. 
If the system caches only latents, the same cache budget covers  more objects, but every cache hit still incurs GPU decode cost.  
The right choice is therefore not a fixed format: newly popular images may be worth keeping as decoded images, while colder images are better kept as compact latents. As image popularity changes over time and new images continuously arrive, the system must adaptively split cache space between images and latents.

We present \system, a novel latent-first storage 
system that stores AI-generated images as compressed latents and reconstructs them on demand using GPU decoding. 
\fref{fig:ls_conceptual} shows the shift from image persistence to latent-first storage, and \fref{fig:intro_scatter} previews the resulting cost--latency tradeoff.  

This paper makes the following contributions:

\begin{itemize}[noitemsep,leftmargin=*]
  \item \textbf{Workload characterization:}
    To the best of our knowledge, we present the first large-scale analysis of how users access AI-generated images , using a 35-month, 2-billion-request production trace from a major generative-content platform. 
  \item \textbf{Latent-first storage:}
    To our knowledge, \system is the first storage system for AI-generated images that treats compressed, model-native latents as durable storage objects and uses on-demand GPU reconstruction on the read path to trade inexpensive compute for large storage savings.  
    \item \textbf{Dual-format adaptive cache:} 
    \system makes latent-first storage practical through a dual-format cache that caches hot objects as decoded images for fast hits and colder objects as compact latents for coverage and an adaptive cache resizer that adjusts the split between above two parts. 
    \item \textbf{End-to-end evaluation:}
    We build a \system prototype and evaluate it using the production trace
    \system reduces persistent storage by 78.7\%, lowers mean and P99 read latency by 17\% and 18\% over PNG-only storage, and projects over 60\% cumulative cost savings over 20 years. 
\end{itemize}

Our implementation and workload trace are publicly available at \url{https://github.com/ds2-lab/LatentStore}.

\section{Background}
\label{sec:background}

\subsection{The Scale of AI-Generated Images}
\label{subsec:bg-scale}

The proliferation of text-to-image generative models has created an unprecedented volume of synthetic visual content.
As shown in \fref{fig:adobe-growth}, Adobe Firefly alone has exhibited explosive growth since its April~2023 launch: the platform reached 3~billion cumulative images within its first six months, crossed 16~billion by the end of 2024, and surpassed 24~billion by mid-2025, with the monthly generation rate accelerating from roughly 500~million to over 2~billion images per month~\cite{adobe2023fireflyfoundry}.
Beyond Adobe, Midjourney produces over 34~million images per day~\cite{everypixel2023aiimagestats}, and open-source platforms built around Stable Diffusion~\cite{rombach2022high, sd3,song2021ddim,peebles2023scalable} and FLUX~\cite{labs2025flux1kontextflowmatching} contribute further to the global supply of AI-generated imagery.

This rapid growth translates directly into a storage challenge. 
At typical resolutions ($1024 \times 1024$), storing 100~billion images as losslessly compressed PNGs requires on the order of 150\,PB, with storage capacity costs scaling linearly with the ever-expanding working set.
Unlike camera photos, irreproducible records of a physical scene, AI-generated images are \emph{deterministic outputs of a computational process} fully characterized by a compact latent tensor, opening an optimization axis that conventional image stores cannot exploit.  

\subsection{Diffusion Models and VAE Decoding}
\label{subsec:bg-diffusion}

\begin{figure}[t]
\centering
\begin{minipage}[t]{0.49\columnwidth}
\centering
\includegraphics[width=\linewidth, height=4.5cm, keepaspectratio]{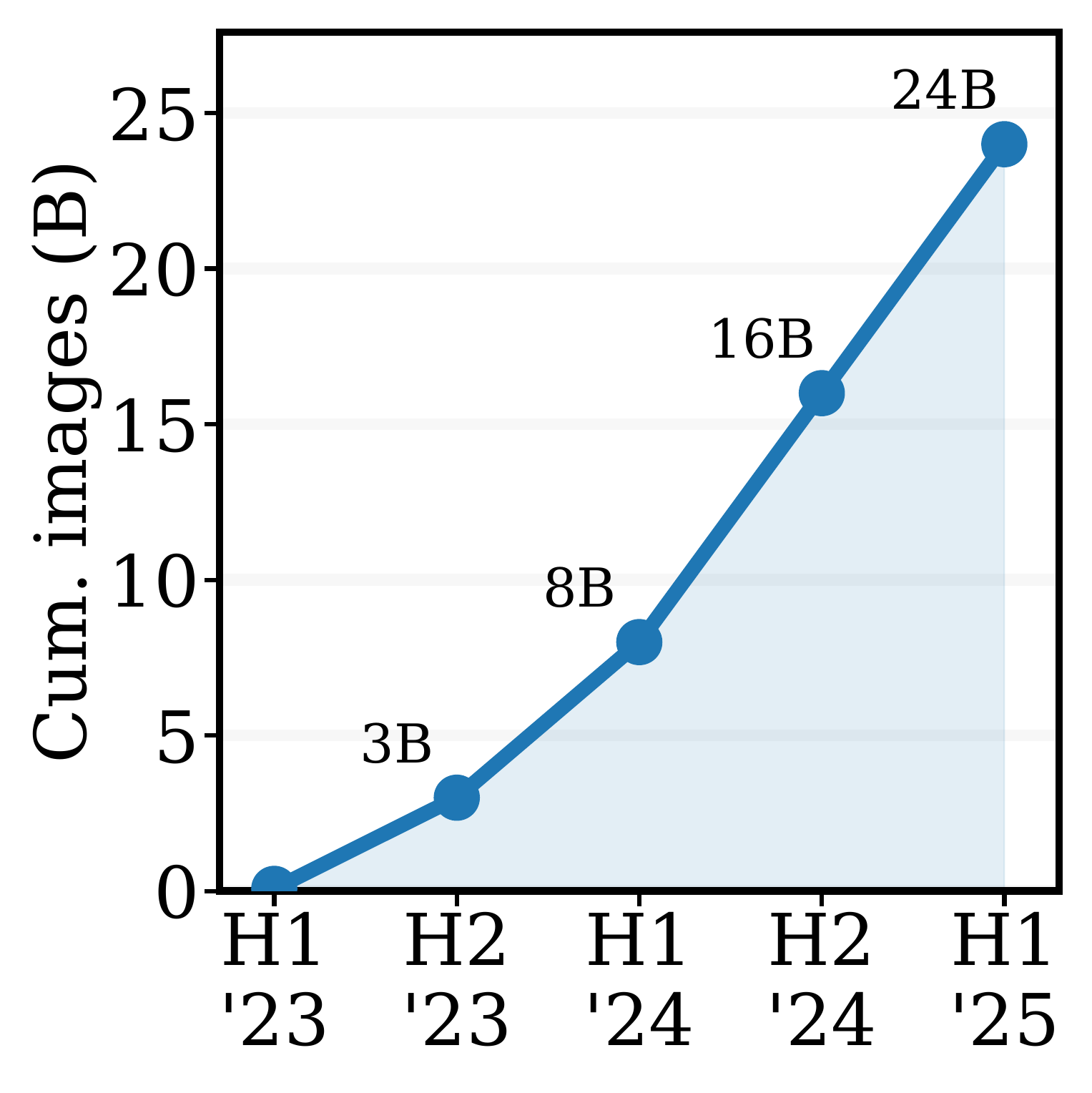}
\vspace{-17pt} 
\captionof{figure}{Adobe Firefly sees explosive gen-image growth~\cite{adobe2023fireflyfoundry}.}
\label{fig:adobe-growth}
\end{minipage}\hfill
\begin{minipage}[t]{0.435\columnwidth}
\centering
\includegraphics[width=\linewidth, height=4.5cm, keepaspectratio]{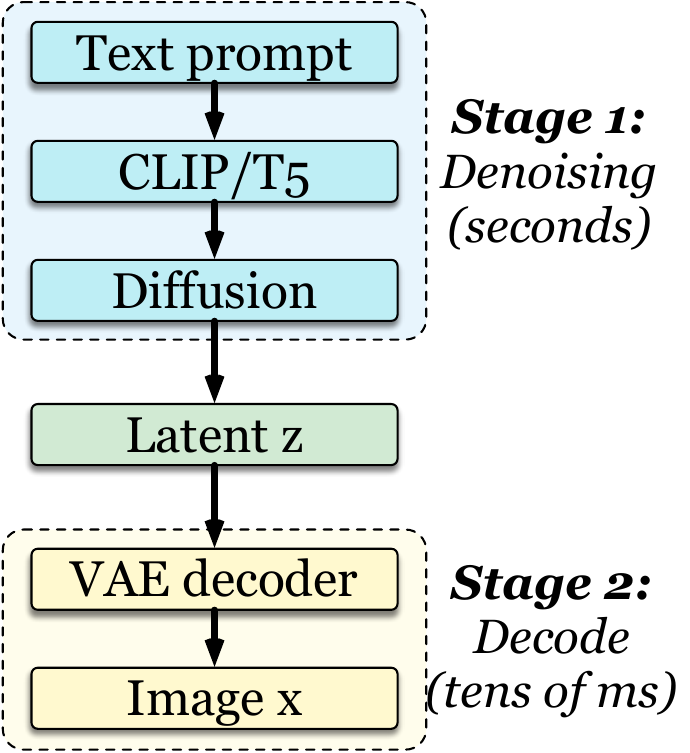}
\vspace{-17pt}
\captionof{figure}{Two-stage text-to-image pipeline.} 
\label{fig:pipeline}
\end{minipage}
\vspace{-3pt}
\end{figure}

\begin{figure*}[t]
  \centering
  \begin{subfigure}[b]{0.24\textwidth}
    \centering
    \includegraphics[scale=0.15]{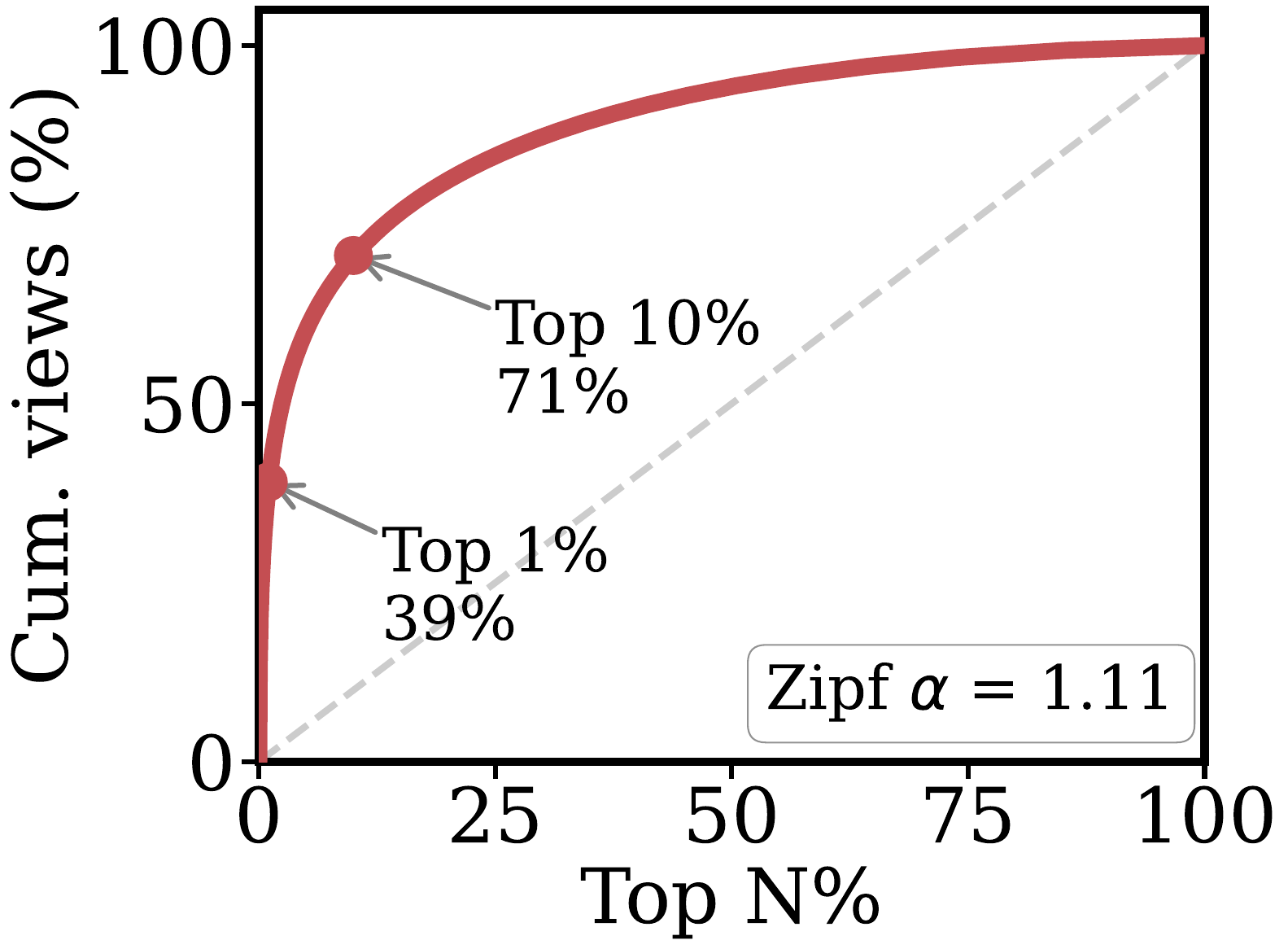}
    \vspace{-5pt}
    \caption{Popularity CDF.}
    \label{fig:trace-pop}
  \end{subfigure}\hfill
  \begin{subfigure}[b]{0.24\textwidth}
    \centering
    \includegraphics[scale=0.15]{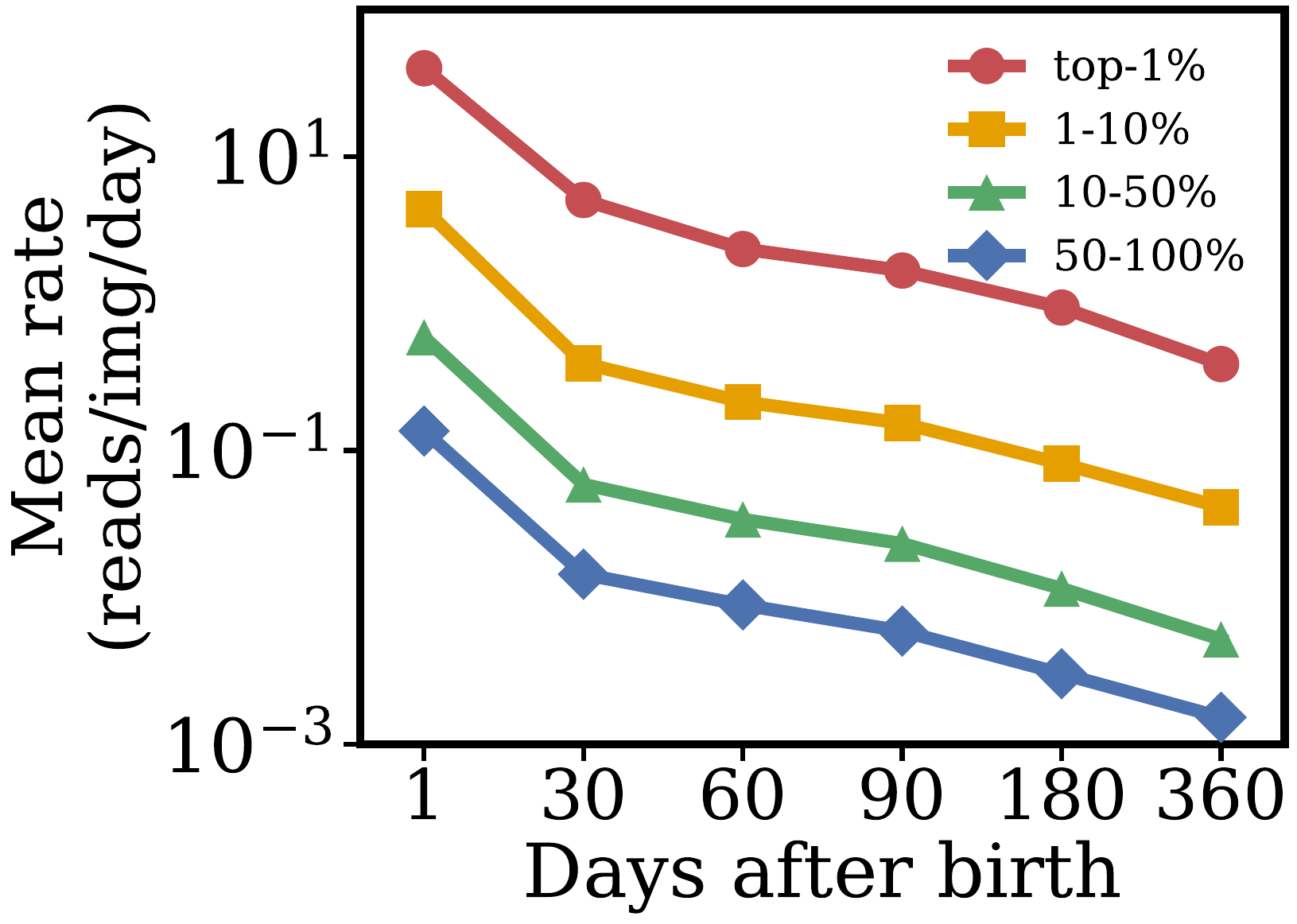}
    \vspace{-5pt}
    \caption{Access decay vs.\ age.}
    \label{fig:trace-age}
  \end{subfigure}\hfill
  \begin{subfigure}[b]{0.24\textwidth}
    \centering
    \includegraphics[scale=0.15]{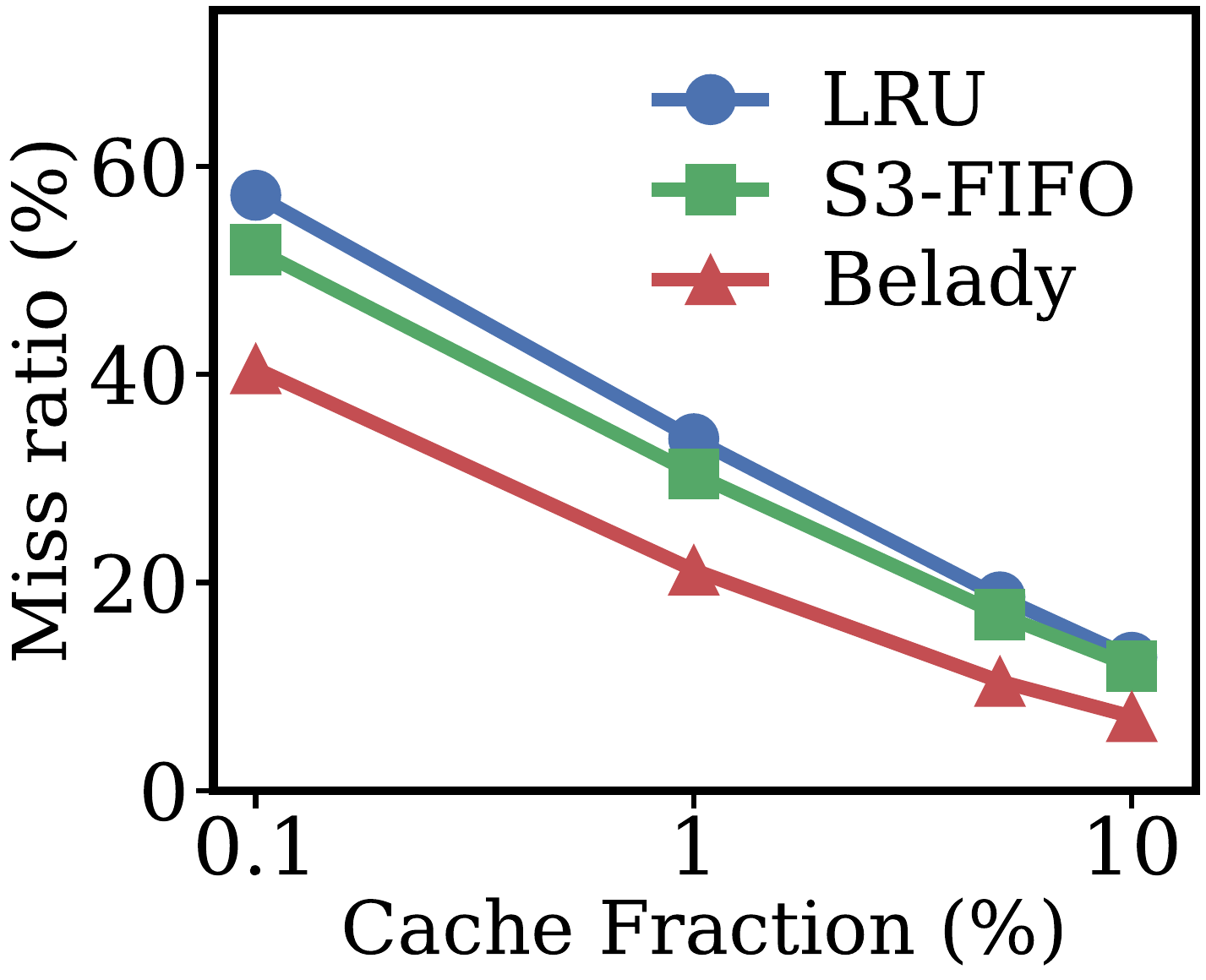}
    \vspace{-5pt}
    \caption{Miss ratio vs.\ cache size.}
    \label{fig:trace-miss}
  \end{subfigure}\hfill
  \begin{subfigure}[b]{0.24\textwidth}
    \centering
    \includegraphics[scale=0.15]{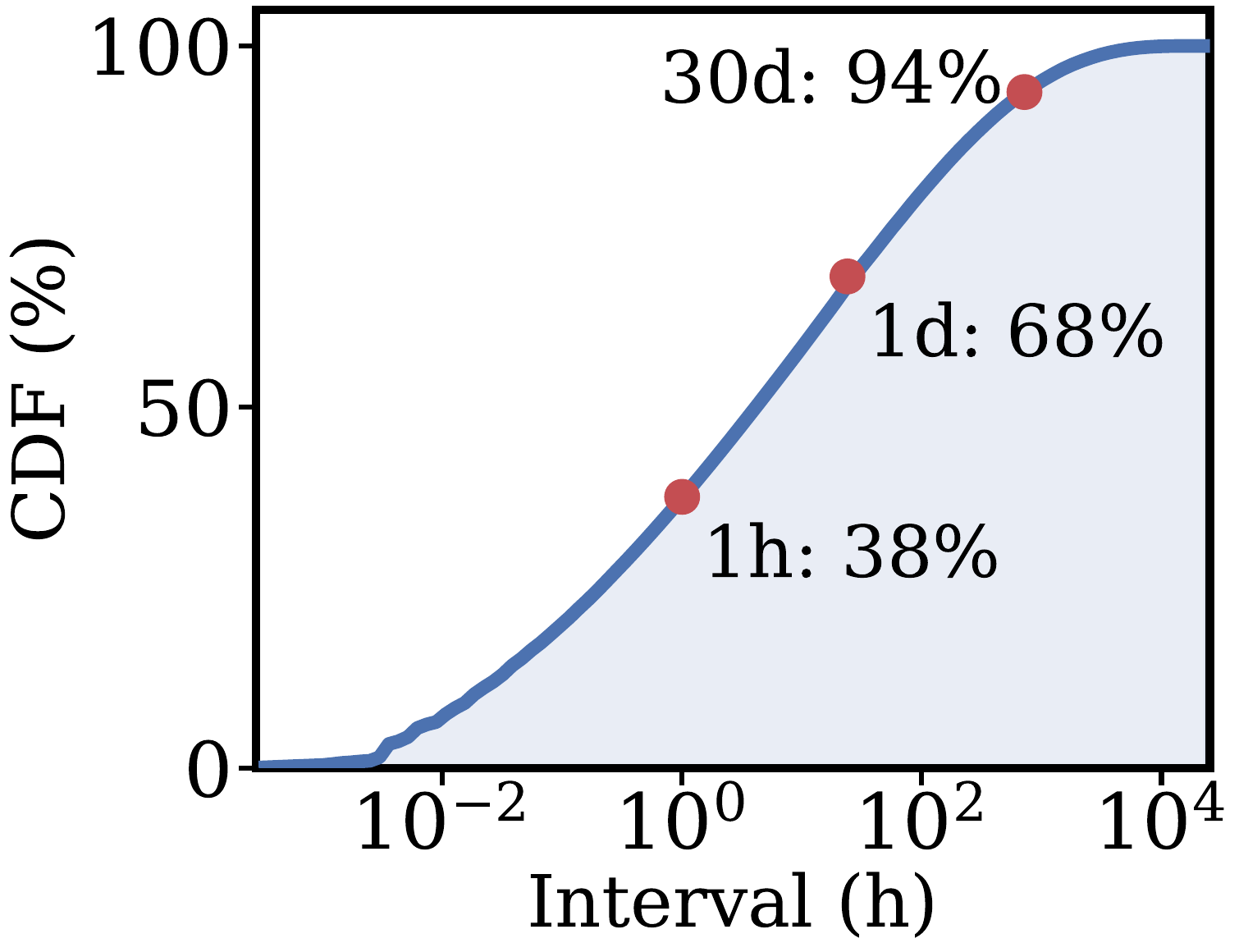}
    \vspace{-5pt}
    \caption{Re-access interval CDF.}
    \label{fig:trace-reaccess}
  \end{subfigure}
  \vspace{-5pt}
  \caption{\civitai trace characterization.
    (a)~Image popularity CDF.
    (b)~Mean access rate vs.\ age, stratified by lifetime-view
    quartile.
    (c)~Miss ratio vs.\ cache size for three policies.
    (d)~CDF of intervals between consecutive accesses.
    }
    \vspace{-12pt}
  \label{fig:trace-char}
\end{figure*}

Modern text-to-image models, including Stable Diffusion and FLUX, 
generate images through a two-stage VAE (variational autoencoder)~ pipeline\cite{kingma2013auto}. See \fref{fig:pipeline}. 

\phead{Stage~1: Denoising.}
A text prompt is first encoded into a conditioning vector by a language model such as CLIP~\cite{radford2021learning,cherti2023reproducible,ramesh2022hierarchical} or T5~\cite{raffel2020exploring}.
A diffusion process then iteratively denoises a random tensor in a \emph{latent space}, guided by the conditioning signal, over 20--50 steps.
The output is a latent tensor $\mathbf{z} \in \mathbb{R}^{c \times h \times w}$, where $c$ is the number of latent channels and $h \times w$ is the spatial resolution of the latent grid.

\phead{Stage~2: VAE decoding.}
The VAE decoder maps the latent tensor to the pixel space: $\mathbf{x} = \mathcal{D}(\mathbf{z})$, producing an RGB image $\mathbf{x} \in \mathbb{R}^{3 \times H \times W}$.
The decoder is a \emph{deterministic} feed-forward neural network with no sampling or stochastic components.
\rev{Throughout this paper we assume a \emph{pinned} decoding configuration (decoder weights, numeric precision, inference engine, and GPU architecture), under which the same latent always yields a bit-identical pixel output.}
This single forward pass is orders of magnitude cheaper than the iterative diffusion process.
Note that while decoding in LLM text generation is memory-bandwidth-bound~\cite{sarathi_serve_osdi24}, denoising and decoding in image generation are compute-bound---the denoising and decoding latency scales linearly with batch size.

\section{Motivation}
\label{sec:motivation}

This section motivates the design of \system by analyzing a production trace (\sref{subsec:mot-trace}) and characterizing the cost of on-demand pixel reconstruction (\sref{subsec:mot-decode}).

\subsection{Production Trace Analysis}
\label{subsec:mot-trace}

To understand the access characteristics of AI-generated images at scale, we collected a 35-month production access trace from \civitai, one of the largest open platforms for generative-AI imagery, where creators publish and share model-generated images and community members browse, download, and remix them. 
The trace records over 2 billion requests from April~2023 to March~2026 and each request log has a \textit{timestamp, image ID, model ID, and model version ID}. 
\tref{tab:trace-overview} summarizes the trace.

\subsubsection{Workload Characterization}
\label{subsec:trace-overview}

The dataset spans 1,049 days and records 2.07~billion anonymized image-view events across 92.3~million unique images generated by more than 710\,K distinct models.
The monthly request volume ranges from 45--70~million views, with an average of roughly 2~million requests per day. 
We make four observations that motivated {\system}.

\phead{Observation \#1 (O1): Skewed popularity, with a heavy tail.}
Image popularity follows a Zipf-like distribution with $\alpha \approx 1.11$
(\fref{fig:trace-pop}).
The top~1\% of images account for 39\% of all views and the top~10\% for 71\%, while 69\% receive fewer than ten views over the trace and 15\% are accessed only once.

\noindent\textbf{\textit{Design implication:}}
This skew suggests that most generated images are rarely accessed after creation, making it inefficient to keep them in hot storage. 
Furthermore, because the images can be reproduced by the model, cold images could be regenerated on demand as long as the model remains available.

\phead{Observation \#2 (O2): Rapid post-birth decay, even for popular images.}
\fref{fig:trace-age} shows the mean per-image access rate as a function of days since first appearance, stratified by lifetime popularity.
Even the most popular 1\% of images see their access rate drop by over $100{\times}$ within a year.
All popularity tiers follow the same pattern: being ``hot'' is a transient \emph{phase}, not a permanent property. Images go from newly uploaded to briefly popular and then quickly become cold, regardless of their eventual total view count.

\noindent\textbf{\textit{Design implication:}} Because popularity is transient rather than intrinsic, the workload mix a serving system faces changes continuously. Any cache format or sizing policy must therefore adapt online: a static configuration becomes increasingly mismatched as content ages and new images arrive.

\phead{Observation \#3 (O3): Non-trivial miss ratios persist even with caching.}
\fref{fig:trace-miss} shows the miss ratio for LRU, S3-FIFO~\cite{s3fifo_sosp23}, and offline-optimal Belady~\cite{belady_ibm66} as the cache grows from 0.1\% to 10\% of the working set.
Even at 10\%, the best online policy (S3-FIFO) still misses roughly 12\% of requests, so a significant fraction always falls through to the slow path.

\noindent\textbf{\textit{Design implication:}} We should size the cache carefully to strike a balance between cost and latency. 

\phead{Observation \#4 (O4): Large variance in re-access intervals.}
\fref{fig:trace-reaccess} shows the CDF of intervals between consecutive accesses to the same image, aggregated over the $\sim$78.1~M images that were accessed more than once. Roughly 38\% of re-accesses happen within an hour and 68\% within a day; the remaining 32\% are spread across days, weeks, and months, with 6\% beyond 30~days. This explains why we still observe over 10\% miss ratio at a large cache size. 

\noindent\textbf{\textit{Design implication:}} The large variance in re-access intervals calls for a multi-tiered cache to reduce both cost and latency.

\begin{table*}[t]
    \centering
    \scriptsize
    \caption{
    (a)~Summary of the 35-month production trace from \civitai.
    (b)~Decoder, latent, and image size across model families at $1024{\times}1024$. Pixel is uncompressed RGB; Lat Compressed is pcodec-compressed latent; PNG is the average file size from our evaluation dataset. SD~1.5 supports only $512{\times}512$.
    (c)~Per-image SD~3.5 latency on NVIDIA GPUs ($1024{\times}1024$, 28 denoising steps, FP16, batch size 1).}
    \label{tab:decode-char}
    \vspace{-5pt}
    \begin{subtable}[b]{0.20\textwidth}
    \centering
    \caption{Trace summary.}
    \label{tab:trace-overview}
    \renewcommand{\arraystretch}{1.15}
    \begin{tabular*}{\linewidth}{@{\extracolsep{\fill}}lr@{}}
    \toprule
    \textbf{Metric} & \textbf{Value} \\
    \midrule
    Requests    & 2.07\,B \\
    Images      & 92.3\,M \\
    Models      & 710\,K \\
    \bottomrule
    \end{tabular*}
    \end{subtable}\hfill
    \begin{subtable}[b]{0.46\textwidth}
    \centering
    \caption{Decoder \& latent size.}
    \label{tab:vae-size}
    \resizebox{\linewidth}{!}{\begin{tabular}{@{}lrrrrrr@{}}
    \toprule
    & & \multicolumn{2}{c}{\textbf{Image}} & \multicolumn{3}{c}{\textbf{Latent}} \\
    \cmidrule(lr){3-4}\cmidrule(l){5-7}
    \textbf{Model} & \textbf{Params} & \textbf{Pixel} & \textbf{PNG} & \textbf{Shape} & \textbf{Size} & \textbf{Compressed} \\
    \midrule
    SD 1.5 (2022) & 49.49\,M & --      & --      & $4{\times}128\times128$  & 128\,KB & --      \\
    \rev{SDXL (2023)}   & \rev{49.5\,M}  & \rev{3.0\,MB} & \rev{1.5\,MB} & \rev{$4{\times}128\times128$}  & \rev{128\,KB} & \rev{83\,KB}  \\
    SD 3.5 (2024) & 49.55\,M & 3.0\,MB & 1.4\,MB & $16{\times}128\times128$ & 512\,KB & 277\,KB \\
    FLUX.1 (2024) & 49.55\,M & 3.0\,MB & 1.3\,MB & $16{\times}128\times128$ & 512\,KB & 321\,KB \\
    \rev{DC-AE (2024)}  & \rev{159.2\,M} & \rev{3.0\,MB} & \rev{1.6\,MB} & \rev{$32{\times}32\times32$}   & \rev{64\,KB}  & \rev{43\,KB}  \\
    \rev{Qwen-Image (2025)} & \rev{73.3\,M} & \rev{3.0\,MB} & \rev{1.4\,MB} & \rev{$16{\times}128\times128$} & \rev{512\,KB} & \rev{296\,KB} \\
    \bottomrule
    \end{tabular}
    }
    \end{subtable}
    \hfill
    \begin{subtable}[b]{0.27\textwidth}
    \centering
    \caption{Per-GPU latency and price~\cite{pcpartpicker2026rtx4090,pcpartpicker2026rtx5090,jarvislabs2026h100price}.}
    \label{tab:gpu-latency}
    \renewcommand{\arraystretch}{1.15}
    \begin{tabular}{@{}lrrr@{}}
    \toprule
    \textbf{GPU} & \textbf{Denoise} & \textbf{Decode} & \textbf{Price} \\
    \midrule
    RTX 4090  & 6.23\,s & 67.2\,ms & ${\sim}\$2$K\\
    RTX 5090  & 4.49\,s & 47.3\,ms & ${\sim}\$3.8$K\\
    H100 PCIe & 4.09\,s & 32.6\,ms & ${\sim}\$25$K\\
    \bottomrule
    \end{tabular}
    \end{subtable}
    \vspace{-8pt}
    \end{table*}

\subsection{Diffusion Model Compute Characteristics}
\label{subsec:mot-decode}

The two-stage computation in the diffusion model presents a unique opportunity to improve storage efficiency---storing the intermediate state (latent) instead of the raw image and reconstructing the images on demand. 

This subsection analyzes the compute characteristics of diffusion models and show that (1) latent is \rev{4--39$\times$} smaller than the corresponding image across different models; (2) decode is around two orders of magnitude faster than generation. 

\phead{Latents are smaller than images.}
\rev{\tref{tab:vae-size} compares latent and image sizes across model families released from 2022 to 2025 at $1024{\times}1024$.
The raw FP16 latent is 64--512\,KB, versus 3.0\,MB of raw pixels; after lossless compression on both sides, a compressed latent is 43--321\,KB, still $4$--$39{\times}$ smaller than the corresponding PNG (1.3--1.6\,MB). For SD~3.5, the ratio is ${\sim}5{\times}$ (277\,KB vs.\ 1.4\,MB).
This gap is unlikely to close as models evolve. The denoiser runs over the latent for tens of steps, so generation cost grows with latent size and model designers keep latents small: from 2022 to 2025 the latent grid never grew ($128{\times}128$, or $32{\times}32$ for DC-AE) and only the channel count rose (4${\to}$16${\to}$32) to improve quality.}

\phead{Decoding is much faster than generation.}
Image generation is typically compute-bound because it requires iterative denoising; for SD~3.5, this stage takes 4--6\,seconds per image, depending on the GPU. In contrast, decoding is a single deterministic forward pass and completes in only tens of milliseconds. \tref{tab:gpu-latency} reports the per-image latency of both stages on three representative GPUs. \emph{Decode latency remains well below the sub-second threshold of interactive image retrieval.}

\phead{Consumer GPUs are cost-effective for decoding.}
The decoding latency gap between datacenter and consumer GPUs is modest. Although RTX~5090 and RTX~4090 are 85\% and 92\% cheaper than H100, decoding a latent on RTX~5090 increases latency by less than 15~ms compared to H100, whereas RTX~4090 increases latency by another 20~ms as shown in \tref{tab:gpu-latency}. By comparison, fetching from AWS S3 takes 100--200 ms~\cite{aws_s3_latency}. 
On-demand reconstruction can thus run efficiently on commodity GPUs, substantially reducing the cost per decode.

\phead{Decoders are lightweight.}
\tref{tab:vae-size} compares decoders across three widely used model families. All \rev{decoders in \tref{tab:vae-size}} are compact neural networks with \rev{49--159}\,M parameters. At bfloat16 precision, each decoder occupies less than \rev{320}\,MB of GPU memory, making it \emph{practical to co-locate multiple decoders on a single GPU or to load a decoder on demand with minimal startup overhead.} \rev{Consequently, a decoder loads onto a GPU with negligible cold-start latency, and multiple decoders can remain co-resident on one GPU.}

In summary, as the volume of generated images continues to grow rapidly, efficient and sustainable storage systems should exploit the distinctive properties of diffusion-based generation: storing less popular or older images as compact latent and reconstructing them only when needed.

\section{{\system} Design}
\label{sec:design}

This section presents the architecture of \system, a distributed serving system that stores AI-generated images as compressed latent tensors and uses a dual-format cache to reduce serving latency and decoding cost. The design of \system must address three coupled challenges.

\phead{Challenge \#1 (C1). }
Although on-demand decoding adds less than 50\,ms to the critical path, it still incurs nontrivial compute overhead, so caching is necessary to avoid repeated decodes of popular content. 
Yet neither single-format extreme suffices. An all-image cache avoids decoding on hits but can cache only a limited number of objects, whereas an all-latent cache\footnote{Caching latents avoids object-store fetch latency (100-200\,ms~\cite{aws_s3_latency}).} maximizes the number of cached items, but forces a GPU decode on every hit. 
Because the workload combines highly skewed popularity (O1) with widely varying re-access intervals (O4), \system must cache both formats at once. 

\phead{Challenge \#2 (C2). }
The optimal fraction of cache capacity devoted to decoded images is workload-dependent: rapid post-birth popularity decay (O2). The continual arrival of new images implies that the value of storing an image vs. a latent depends on the popularity and how long they are cached. 

\phead{Challenge \#3 (C3). }
Routing must simultaneously preserve cache locality and balance GPU load. Least-loaded routing spreads requests for the same image across nodes, leading to redundant caching of hot content and reducing effective aggregate capacity, while imbalanced load for latent wastes GPU cycles. 
\system therefore requires a unified design that jointly manages the cache and request routing. 

\system addresses C1 with a \emph{dual-format cache}
(\sref{sec:dual-format}), C2 with \emph{online marginal-hit tuning} of the image-to-latent ratio (\sref{sec:autotune}), and C3 with \emph{consistent-hashing routing} that uses a cache-pinned spillover (\sref{sec:routing}).

\subsection{Design Overview} 
\label{sec:arch-overview}

\phead{Latent representation as a storage object.}
The denoising process maps a text prompt
to a compact latent tensor, which lossless compression shrinks further, so a stored compressed latent is much smaller than the decoded image. 

\begin{figure}[t]
  \centering
  \includegraphics[width=0.48\textwidth]{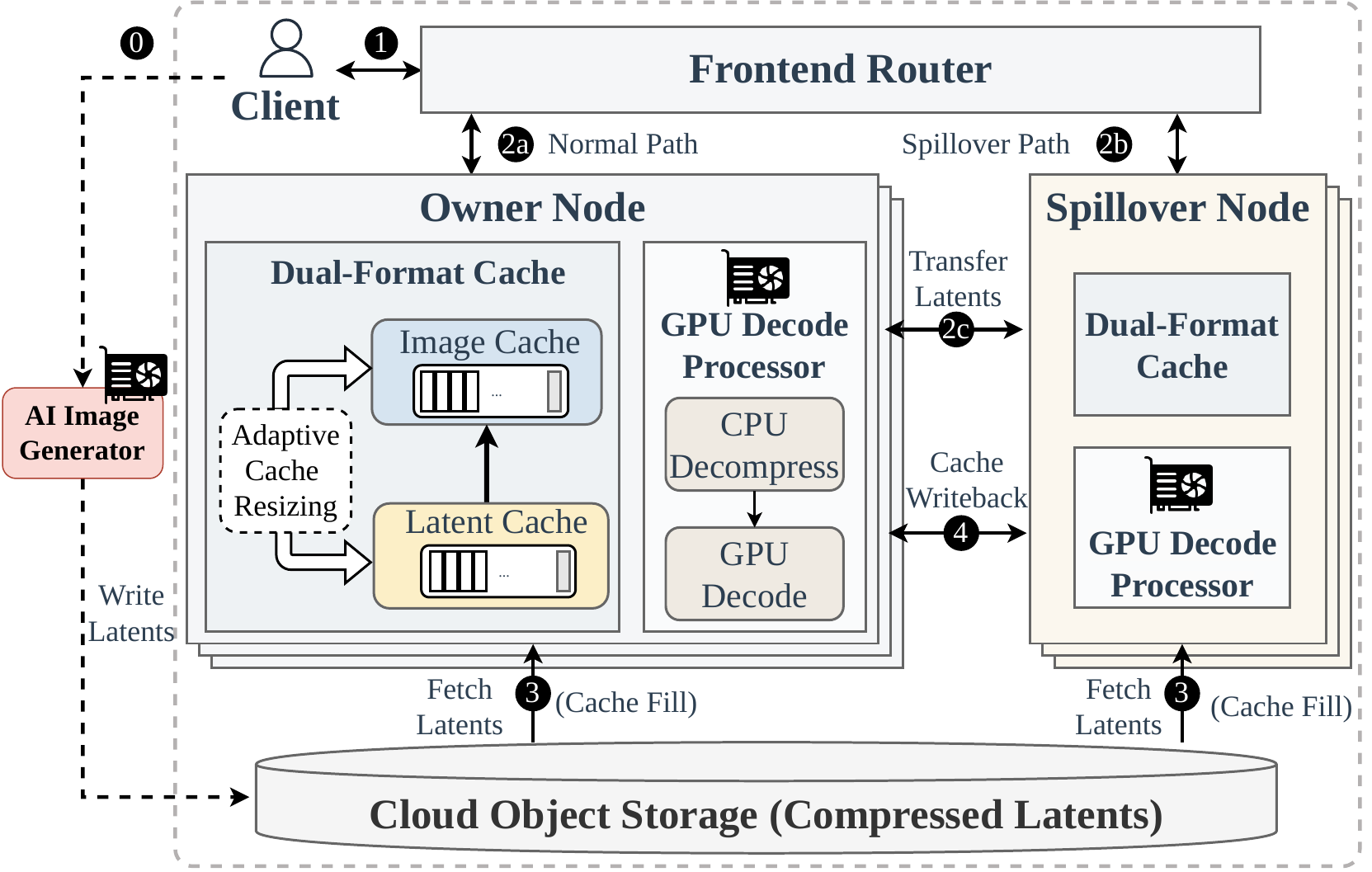}
  \vspace{-15pt}
  \caption{\rev{\system architecture and request flow.}}
  \label{fig:arch}
  \vspace{-3pt} 
\end{figure}

\phead{System components.}
\fref{fig:arch} shows the architecture of {\system}. 
A \emph{frontend router} fronts a fleet of homogeneous GPU \emph{nodes}, backed by a cloud object store (e.g., Amazon~S3) that durably persists the compressed latents. The router maintains two pieces of lightweight runtime state: a map of in-flight image identifiers, so that a burst of identical requests issues only one decode, and a \emph{routing table} that uses consistent hashing to map each identifier to its \emph{owner node} and tracks per-GPU queue depths for load-aware dispatch. Every GPU node is functionally identical: it hosts a \emph{dual-format cache} that holds each object either as a decoded image or as a compressed latent (\sref{sec:dual-format}), an \emph{adaptive resizer} that balances the two tiers online (\sref{sec:autotune}), and one \emph{decode pipeline} per GPU streaming CPU decompression and GPU decode (\sref{sec:implementation}).  

\phead{Request flow.}
\rev{Image generation is handled by a separate generation system (\circled{0} in \fref{fig:arch}): when a client issues a generation request, that system produces the image and writes only its compressed latent to the object store, and \system serves all subsequent reads. A read request (\circled{1})} first reaches the router, which checks for an in-flight decode and uses consistent hashing to map it to its owner node. Two paths follow:
\begin{itemize}[noitemsep,leftmargin=*]
  \item \textbf{Normal path} (\circled{2a}):
    The owner node is not overloaded, so the router dispatches directly. The node probes its cache: an image-tier hit returns immediately; a latent-tier hit triggers local GPU decode; a full miss fetches the compressed latent from cloud storage (\circled{3}) before decoding.
  \item \textbf{Spillover path} (\circled{2b}--\circled{4}): 
    the owner node's GPU queue exceeds a threshold $\theta$, so the router redirects the request to a less-loaded
    \emph{spillover node}. If the owner already holds the latent, it is shipped to the spillover node (\circled{2c}, ``share latents'');  otherwise the spillover node fetches it from cloud storage (\circled{3}). After GPU decoding, the result is written back to the owner node's cache (\circled{4}), so the entry stays where consistent hashing placed it.
\end{itemize}
The cloud object store thus serves as the single source of truth for every object and is touched only on full cache misses; latents average about 20\% the size of decoded images, dramatically reducing both storage footprint and network transfer cost versus persisting full-resolution pixels (\sref{subsec:storage}).
\subsection{Dual-Format Cache}
\label{sec:dual-format}
Caching decoded images reduces serving latency, but limited cache capacity restricts how many images can be stored. Caching compact latents fits many more items (more coverage) but requires GPU decoding. One option is to keep both formats in a single mixed-format LRU. However, a given capacity cut-off contains an unpredictable mix of the two formats whose composition shifts with every access, making it infeasible to control the memory partition precisely.

\system therefore maintains a \emph{dual-format cache} on each node: two independent cache tiers, one for decoded images and one for compressed latents. The two caches share a fixed total capacity $C$. An adaptive cache resizer assigns an $\alpha$ fraction of $C$ to the \emph{image cache} and the remaining $1{-}\alpha$ to the \emph{latent cache}. In this way, each cache's capacity is fully determined by $\alpha$, and each cache's miss ratio curve reflects a single, homogeneous object format, which is the prerequisite for the marginal tuning 
in \sref{sec:autotune}. 
\fref{fig:cache} illustrates the design. 
Table~\ref{tab:dualformat-vars} summarizes the notation used in this section.

\phead{Lookup flow}
The two caches have two operational characteristics.
First, lookup is cascading: image cache first, then latent cache, and finally a cloud fetch on a full miss.
Second, every object lives in exactly one cache at a time. A new object enters the latent cache on its first cloud fetch with a hit counter at zero; each subsequent latent-cache access increments the counter; once the counter crosses a promotion threshold $h$, the object is decoded, inserted into the image cache, and atomically removed from the latent cache. 

\phead{Per-access latency costs.}
Every lookup resolves to one of three outcomes with sharply different costs. An image-cache hit returns the decoded image immediately and adds no latency. A latent-cache hit incurs $T_{decode}$, the GPU decode latency needed to reconstruct pixels from the compressed latent. A full miss incurs a cloud object-store fetch of the latent, with a cost at $T_{fetch}$, plus the same decode latency $T_{decode}$(\fref{fig:cache}).

\subsection{Online Marginal-Hit Tuning}
\label{sec:autotune}

To address Challenge 2, \system needs a mechanism to tune the image-to-latent ratio $\alpha$ online.
A natural starting point is the \emph{miss-ratio curve} (MRC)~\cite{mattson_ibm70, counterstacks_osdi14, shards_fast15, cliffhanger_nsdi16}: given a cache of capacity $C$, the MRC maps $C$ to the miss ratio under a particular replacement policy. If the MRC were available for both caches, one could sweep over $\alpha$ and pick the split that minimizes 
overall latency. 
However, constructing full MRCs is impractical online. Moreover, the dual-format cache creates a \emph{dependency} between the two caches: the latent cache sees only requests that missed the image cache, so its MRC changes whenever $\alpha$ changes. Globally minimizing latency would require the entire family of latent-cache MRCs indexed by image-tier size ($O(C^2)$), which is prohibitively expensive to maintain online via shadow caches or trace replay. 

\system sidesteps global MRC construction entirely. Instead of asking ``what is the miss ratio at every possible capacity?'', it asks a strictly local, \emph{marginal-hit} question at each tuning window: \emph{which tier benefits more from one additional unit of capacity at the current operating point?}

\begin{figure}[t]
  \centering
  \includegraphics[width=0.42\textwidth]{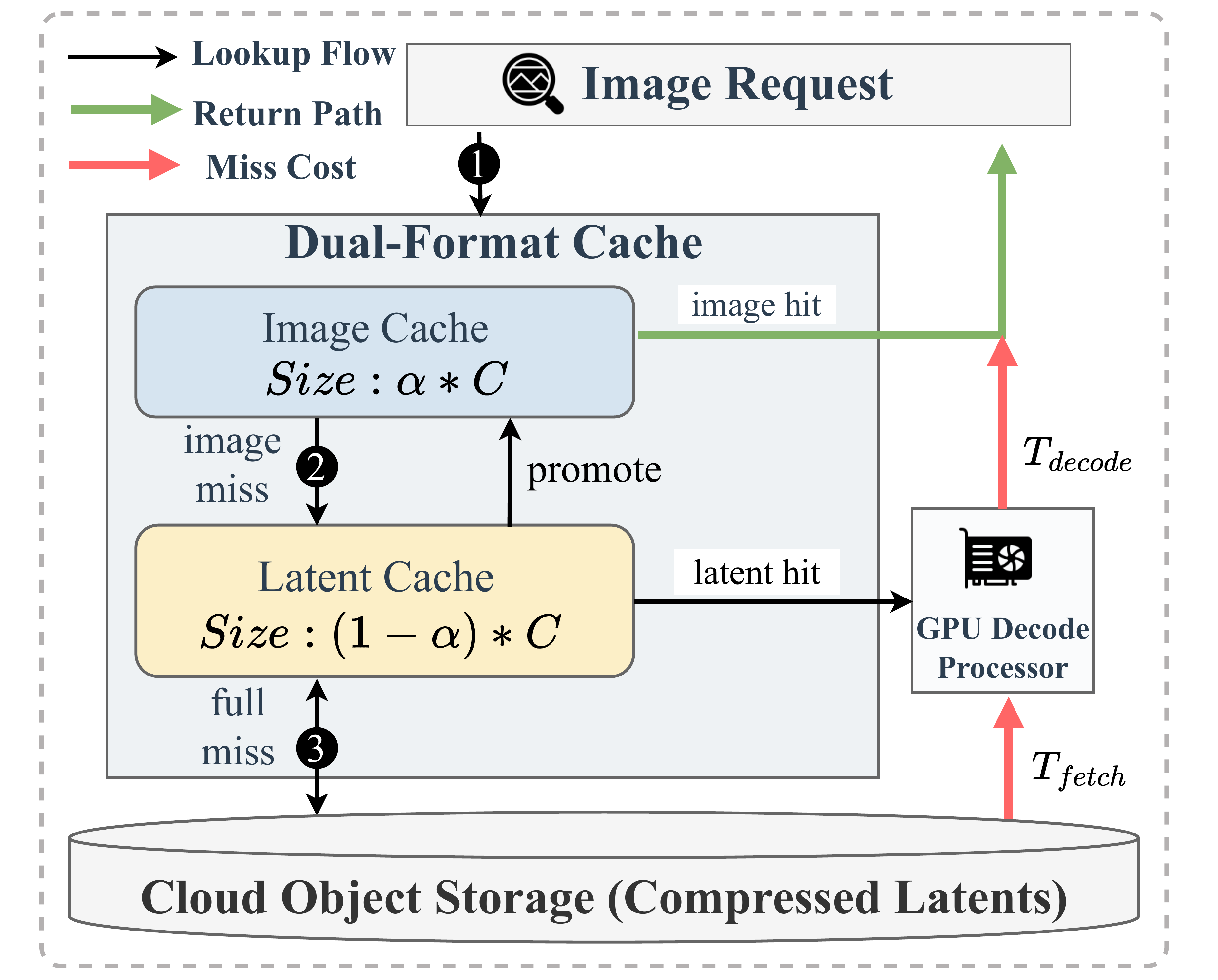}
  \vspace{-5pt} 
  \caption{Dual-format cache design.} 
  \label{fig:cache}
  \vspace{-3pt}  
\end{figure}

\phead{Cost model and gradient-based tuning.}
Let $\mathrm{MR}_{\text{img}}(\alpha)$ denote the image-cache miss ratio at this allocation, measured over the full request stream: a request counts as an \emph{image miss} if the requested object is not found in the image cache, regardless of whether it is later found in the latent cache.
Let $\mathrm{MR}_{\text{lat}}(\alpha)$ denote the latent-cache miss ratio at this allocation, measured over the \emph{image-miss stream}, i.e., only requests that already missed the image cache. A latent miss is thus a \emph{full miss}: the object is absent from both caches and must be fetched from the cloud. With cascading lookup (image cache $\to$ latent cache $\to$ cloud), the expected per-request cost is:
\begin{equation}
\vspace{-5pt}
\scalebox{0.82}{$\displaystyle
\begin{aligned}
  & E[T](\alpha) = \underbrace{\bigl(1-\mathrm{MR}_{\text{img}}(\alpha)\bigr)\cdot 0}_{\text{image-cache hit}} \\
  & \quad + \mathrm{MR}_{\text{img}}(\alpha)\cdot\Bigl[\underbrace{\bigl(1-\mathrm{MR}_{\text{lat}}(\alpha)\bigr)\cdot T_{decode}}_{\text{latent-cache hit}} + \underbrace{\mathrm{MR}_{\text{lat}}(\alpha)\cdot\bigl(T_{decode}+T_{fetch}\bigr)}_{\text{full miss}}\Bigr]
\end{aligned}$}
\label{eq:latency}
\end{equation}

Differentiating \eref{eq:latency} with respect to $\alpha$ at the current operating point gives a scalar gradient $D$ whose sign directly prescribes the update direction:
\vspace{-5pt}
\begin{equation}
\begin{split}
  \rev{D = {}} & \rev{-\delta_{img}\, T_{decode} + T_{fetch}\cdot S,} \\
  \rev{S = {}} & \rev{\mathrm{MR}_{\text{img}}(\alpha)\,\delta_{lat} - \delta_{img}\,\mathrm{MR}_{\text{lat}}(\alpha),}
\end{split}
  \label{eq:gradient}
\vspace{-7pt}
\end{equation}
where $\delta_{img}$ measures how many additional image misses would appear if the image tier shrank by a small amount, and $\delta_{lat}$ measures how many additional full misses would appear if the latent tier were slightly smaller.
\rev{The term $-\delta_{img}T_{decode}$ is the decode cost of the extra image misses that a smaller image tier would add, and $S=\partial(\mathrm{MR}_{\text{img}}\mathrm{MR}_{\text{lat}})/\partial\alpha$ is how much the full-miss ratio, and hence the number of cloud fetches, grows when the image tier expands at the latent tier's expense.}

If $D < 0$, the image tier has higher marginal value, so $\alpha$ increases by a fixed step $\Delta$; if $D > 0$, the latent tier does, so $\alpha$ decreases by $\Delta$. No global sweep is needed: the sign of $D$ at the current operating point suffices for one improving step.

\label{sec:hybrid-cache}

\begin{table}[t]
\footnotesize
\centering
\caption{Variables of the dual-format cache and marginal-hit tuning.}
\vspace{-5pt}
\setlength{\tabcolsep}{4pt}
\begin{tabular}{@{}ll@{}}
\toprule
Symbol & Definition \\
\midrule
\multicolumn{2}{@{}l}{\emph{Dual-format cache (\sref{sec:dual-format}):}} \\
$C$          & Per-node total cache capacity (bytes). \\
$\alpha$     & Fraction of $C$ given to the image cache; $1{-}\alpha$ to the latent cache. \\
$T_{decode}$        & GPU decode latency. \\
$T_{fetch}$        & Cloud fetch latency. \\
$h$          & Latent-to-image promotion threshold (hit counter). \\
\midrule
\multicolumn{2}{@{}l}{\emph{Online marginal-hit tuning (\sref{sec:autotune}):}} \\
$E[T](\alpha)$ & Expected per-request latency at allocation $\alpha$. \\
$\mathrm{MR}_{\text{img}}(\alpha)$ & Image-cache miss ratio over all requests. \\
$\mathrm{MR}_{\text{lat}}(\alpha)$ & Latent-cache miss ratio, conditional on image-cache miss. \\
$\delta_{img}$   & Marginal image-miss rate (image-cache tail-hit fraction). \\
$\delta_{lat}$   & Marginal full-miss rate (latent-cache tail-hit fraction). \\
$D$          & Scalar gradient $\partial E[T]/\partial \alpha$; sign sets update direction. \\
$W$          & Tuning window size (number of requests). \\
$\Delta$     & Per-window step size applied to $\alpha$. \\
$\tau$       & Fraction of each cache tier reserved as the tail segment. \\
\bottomrule
\end{tabular}
\label{tab:dualformat-vars}
\end{table}

\phead{Online marginal measurement and cost update.}
$\mathrm{MR}_I$ and $\mathrm{MR}_L$ are trivially counted along the lookup path. The marginal rates $\delta_{img}$ and $\delta_{lat}$ require slightly more structure. \system splits each tier into a \emph{main} segment of fraction $1{-}\tau$ and a thin \emph{tail} segment of fraction $\tau$. Items evicted from the main enter the tail; items evicted from the tail leave the cache. A \emph{tail hit} (a request served from the tail rather than the main segment) identifies a request that would have been a miss had that tier been $\tau$ smaller, so the tail-hit rate directly measures the marginal value of the last $\tau$ fraction of capacity.
Concretely, all four statistics are accumulated over a window of $W$ requests, with hit, miss, and total-request counts all measured under the current partition at $\alpha$.
\begin{equation*}
\scalebox{0.88}{$\displaystyle
  \mathrm{MR}_{img}(\alpha) = \frac{\text{image misses}}
                       {\text{total requests}},\qquad
  \delta_{img}(\alpha)
    = \frac{\text{image tail hits}}
           {\text{total requests}},
$}
\end{equation*}
\vspace{-6pt}
\begin{equation*}
\scalebox{0.88}{$\displaystyle
  \mathrm{MR}_{lat}(\alpha) = \frac{\text{full misses}}
                       {\text{image misses}},\qquad
  \delta_{lat}(\alpha)
    = \frac{\text{latent tail hits}}
           {\text{image misses}}.
$}
\end{equation*}
After each window the gradient $D$ is re-evaluated, $\alpha$ is updated, tier capacities are recomputed, overflows are evicted, and all counters reset.
This procedure adds \emph{near-zero overhead}: the tail is carved from the existing cache budget without extra memory, and the four counters are maintained as side effects of the normal lookup path. No global MRC is ever reconstructed; the allocator moves $\alpha$ incrementally using only the gradient observed at the current operating point.

$T_{decode}$ and $T_{fetch}$ are tracked as exponentially weighted moving averages of observed GPU-decode and cloud-storage-fetch latencies. This closes a negative feedback loop that absorbs both workload shifts and transient infrastructure events with a single mechanism (\fref{fig:cache}). When the GPU is overloaded, $T_{decode}$ rises and the conditional cost of an image miss grows; the gradient pushes $\alpha$ upward to enlarge the image cache and offload GPU work.
\rev{Conversely, when storage latency spikes, $T_{fetch}$ rises. Since $T_{fetch}$ scales only $S$ in \eref{eq:gradient}, a rising $T_{fetch}$ shifts capacity toward the latent cache whenever $S>0$, i.e., whenever a larger image tier would cause more full misses. In our experiments $S$ is positive in practice, because compressed latents pack ${\sim}5{\times}$ more objects per byte, so shrinking the latent tier loses far more coverage than the image tier gains; in the rare windows where the latent tail prevents no fetches ($\delta_{lat}{\approx}0$), $S$ is no longer positive and the gradient correctly stops favoring the latent tier.} Popularity drift, GPU throttling, and storage back-pressure thus all feed into the same tuning loop without separate mechanisms.

\subsection{Routing with Spillover} 
\label{sec:routing}

To address C3, \system chains three router-side stages that decouple \emph{where the decode runs} from \emph{where the cache entry lives}: \emph{request coalescing} absorbs duplicate bursts, \emph{consistent-hashing dispatch} pins each item to a stable owner node, and \emph{spillover with cache pinning} reconciles locality with GPU load by letting work move between nodes while every cache entry stays anchored at its hash-determined home.

\phead{Request Coalescing.}
The router maintains a map from image identifiers to in-flight decode requests. A new request for an in-flight identifier waits on the existing decode rather than issuing a duplicate. 
For bursty workloads where the same image is requested hundreds of times per second, coalescing substantially reduces the effective GPU decode load. 

\phead{Consistent-hashing routing.}
The router uses consistent hashing to map each identifier to its owner node and, within that node, dispatches to the least-loaded GPU based on queue depths reported back to the router. Repeated accesses to the same item therefore land on the same owner node and reuse the same cached entry, eliminating cross-node duplication and maximizing aggregate cache effectiveness.

\phead{Spillover with cache pinning.}
When the least-loaded GPU on the owner node exceeds a queue-depth threshold $\theta$, the router activates the spillover path (\circled{\footnotesize{2b}}--\circled{4} in \fref{fig:arch}). The central challenge is that the GPU work must move to a less-loaded node while the cache entry must remain on the hash-determined Owner Node.  
\system resolves this by forwarding the decode task (along with cached latent if any) to the globally least-loaded spillover node and, once the decode completes, writing the result back to the owner node's cache.  

This design guarantees two properties. First, \emph{cache coherence}: every item's authoritative cache entry lives on exactly one node, regardless of which GPU did the decode. Second, \emph{load spillover}: under extreme load \system borrows GPU capacity from less-loaded nodes while preserving cache locality, without dropping requests or fragmenting cache state.

\section{Implementation}
\label{sec:implementation}

We implemented \system in Python atop Ray~\cite{moritz2018ray}
for distributed actors and TensorRT~\cite{tensorrt} for GPU decoding.

\phead{Distributed actor placement.}
\system maps its logical architecture onto Ray actors with node-affinity scheduling. Each physical GPU node hosts one \emph{cache actor} that manages the node's dual-format cache, and one \emph{decoder actor} per GPU. Placement constraints co-locate a node's actors on one machine, so cache lookups and latent transfers are local memory operations. The frontend router runs as an asynchronous event loop, 
issuing non-blocking remote calls to per-node
actors. 

\phead{Pipelined GPU decoding.}
\label{sec:pipeline}
\system decomposes each decode into a multi-stage pipeline spanning three specialized thread pools: an \emph{I/O pool} for cloud storage reads, a \emph{compute pool} for CPU-bound decompression, host-to-device transfer, and image encoding, and a \emph{GPU pool} that serializes TensorRT inference behind an asynchronous lock to ensure mutual exclusion on the device. 
This pipeline keeps the GPU busy while other requests fetch, decompress, or encode, thereby improving throughput under concurrent load.

\phead{TensorRT and CUDA Graphs.}
The decoder is ahead-of-time compiled to a TensorRT engine with FP16 precision and a fixed input shape.
At build time, \system exports the PyTorch model to ONNX~\cite{onnx}, applies TensorRT's layer-fusion and kernel-auto-tuning passes, and caches the resulting engine plan to disk for fast subsequent loads.
At runtime, the engine is wrapped in a CUDA Graph~\cite{cuda_graphs} that captures the entire decode kernel sequence into a single launchable unit, eliminating per-decode kernel dispatch overhead and reducing decode latency compared to eager execution.

\phead{Latent compression.}
Compressed latents stored in cloud object storage are encoded with pcodec~\cite{pcodec_arxiv25}, a lossless compressor designed for columnar numeric arrays.
We choose pcodec over general-purpose byte-stream compressors (e.g., zstd~\cite{zstd}, LZ4~\cite{lz4}) because diffusion-model latents are floating-point tensors with high spatial smoothness and inter-channel correlation, which byte-oriented entropy coders cannot exploit effectively.
The resulting compact representation reduces both storage cost and network transfer time on cache misses.

\section{Evaluation}
\label{sec:evaluation}

Our evaluation answers \rev{six} questions:
\begin{itemize}[noitemsep,leftmargin=*]
  \item \textbf{Q1:} What \emph{data reduction ratio} does \system achieve relative to storing images? (\sref{subsec:storage})
  \item \textbf{Q2:} Does \system sacrifice read latency? (\sref{subsec:e2e})
  
  \item \textbf{Q3:} What \emph{long-term cost savings} does \system provide over a multi-year horizon compared to image storage, and how does the picture shift across GPU price points? (\sref{subsec:cost-projection})

  \item \rev{\textbf{Q4:} How does \system compare with alternative designs in storage and compute cost? (\sref{subsec:spectrum})}

  \item \textbf{\rev{Q5}:} How important is each design choice, including dual-format caching, online tuning, and spillover dispatch, and how sensitive is \system to parameter settings? (\sref{subsec:ablation})

  \item \textbf{\rev{Q6}:} How does \system compare with lossy image compression? (\sref{subsec:fidelity})
\end{itemize}
\rev{Throughout this section, italics highlight each experiment's hypothesis and takeaway.}
\subsection{Experimental Setup}
\label{subsec:setup}

\phead{Testbed.}
We deploy \system on a cluster of three homogeneous GPU nodes, each equipped with one NVIDIA H100 80\,GB PCIe GPU, an AMD EPYC 9554 CPU (28 vCPUs), 177\,GB host memory, and 25\,Gbps network connectivity.
We use AWS~S3 (us-east-1) as the backend storage and an EC2 c6in.4xlarge instance in the same region as the client.
Each GPU node is allocated a 2\,GB dual-format cache \rev{(6\,GB in total, ${\sim}$3\% of the unique-object footprint)}.
We chose H100 GPUs for their wide availability on cloud platforms; consumer GPUs (e.g., RTX~5090) were unavailable for rent on major cloud providers at the time of our experiments.

\phead{Dataset.}
We generate 150\,K images at $1024{\times}1024$ resolution using Stable Diffusion~3.5~\cite{sd3} and store both the pixel PNGs  and pcodec-compressed latents in S3. 
The full dataset occupies 210.6\,GB as PNG and 41.5\,GB as compressed latents.
To derive a replay workload from the \civitai trace (\sref{subsec:trace-overview}), we randomly sample 150\,K unique object IDs (${\sim}615{\times}$ object-level downsample), preserving all accesses to the sampled objects, and map each to one of our generated SD~3.5 images.
The resulting \emph{downsampled trace} contains 57.2\,M requests and is used for sensitivity analysis in \sref{subsec:ablation}.

For the end-to-end latency evaluation (\sref{subsec:e2e}), we select a contiguous 48-hour window from the downsampled trace and replay it at $10{\times}$ wall-clock speed, preserving the original request ordering.
This \emph{evaluation window} contains 66{,}581 requests targeting 12{,}328 unique objects whose aggregate PNG footprint is 21.4\,GB (3.6\,GB as compressed latents).
We pre-warm each node's cache with the preceding requests.

\phead{Baselines and ablations.}
We evaluate \rev{seven} configurations: on-demand generation, ImgStore, and \system variants.

\noindent\emph{Upper-bound reference:}
\begin{itemize}[noitemsep,leftmargin=*]
  \item \textbf{On-Demand Generation}: every previously unseen request triggers the full SD~3.5 diffusion pipeline with 28 denoising steps on a GPU.
    A local image cache serves repeat reads without re-generation.
    This configuration establishes the cost of on-demand generation and is evaluated on a 1{,}000-request subset using the same 3-node GPU cluster.
\end{itemize}
\noindent\emph{Store-and-read configurations:}
\begin{enumerate}[noitemsep,leftmargin=*]
  \item \textbf{Decode-All}: every request fetches a compressed latent from S3 and performs GPU decode. No local caching is used, establishing the raw read latency floor.
  \item \textbf{ImgStore}: a distributed image server backed by S3-stored PNG images with a 2\,GB per-node LRU cache. Cache hits return pre-decoded bytes with no GPU involvement; misses fetch the full PNG from S3.
  \item \textbf{LS-ImgCache} ($\alpha{=}1.0$): \system with an image-only cache; the entire 
  cache budget stores decoded images. Misses fetch and decode compressed latents from S3.
  \item \textbf{LS-LatentCache} ($\alpha{=}0.0$): \system with a latent-only cache; the cache stores only compressed latents. Every cache hit still requires GPU decode.
  \item \textbf{LS-Adaptive}: \system with the dual-format cache and marginal-hit online tuning, adaptively splitting each node's budget between an image tier and a latent tier.
  \item \rev{\textbf{LS-S3ImgCache}: a near-stateless variant that caches decoded images of a hot set in S3; hits bypass the GPU, and misses are decoded and asynchronously added to S3.}
\end{enumerate}

\phead{Metrics.}
For storage, we report the \emph{data reduction ratio} (DRR), defined as the fraction of PNG bytes eliminated: 
$\text{DRR} = (S_{\text{PNG}} - S_{\text{comp}}) / S_{\text{PNG}}$, where $S_{\text{comp}}$ is the compressed-latent footprint under \system.
For 
read access, we report mean, P50, P95, and P99 \emph{read latency}, defined as the wall-clock time from request arrival at the frontend router to response completion.
This metric covers the entire in-cluster processing pipeline, including cache lookup, S3 fetch (on miss), GPU decode, GPU queuing, and data transfer back to the router, but excludes the variable client-to-cluster network hop.
For cache-level analysis, we also report image-tier hit rate, latent-tier hit rate, and full-miss rate.

\begin{table}[t]
\centering
\caption{Storage footprint by model and resolution. ImgStore: baseline; LS: \system (full pcodec-compressed format).}
\label{tab:storage}
\vspace{-5pt}
\scalebox{0.78}{\begin{tabular}{@{}llrrrc@{}}
\toprule
\textbf{Model} & \textbf{Res.} & \textbf{\#Imgs} & \textbf{ImgStore} & \textbf{LS} & \textbf{DRR (\%)} \\
\midrule
SD 3.5 & $1024\times1024$ & 150\,K & 210.6\,GB & 41.5\,GB & 80.3 \\
SD 3.5 & $512\times512$  & 150\,K &  57.1\,GB & 10.9\,GB & 80.8 \\
FLUX.1 & $1024\times1024$ & 100\,K & 130.4\,GB & 32.1\,GB & 75.4 \\
FLUX.1 & $512\times512$  & 100\,K &  35.9\,GB &  8.0\,GB & 77.6 \\
\midrule
\multicolumn{2}{@{}l}{\textbf{Total}} & \textbf{500\,K} & \textbf{434.1\,GB} & \textbf{92.6\,GB} & \textbf{78.7} \\
\bottomrule
\end{tabular}}
\end{table}

\begin{figure*}[t]
\centering
\begin{subfigure}[t]{0.24\textwidth}
\centering
\includegraphics[width=\linewidth]{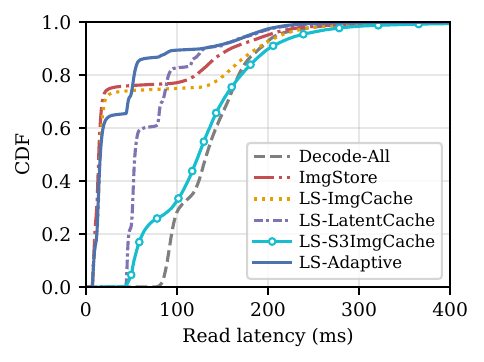}
\caption{\rev{Read latency CDF.}}\label{fig:latency-cdf}
\end{subfigure}\hfill
\begin{subfigure}[t]{0.24\textwidth}
\centering
\includegraphics[width=\linewidth]{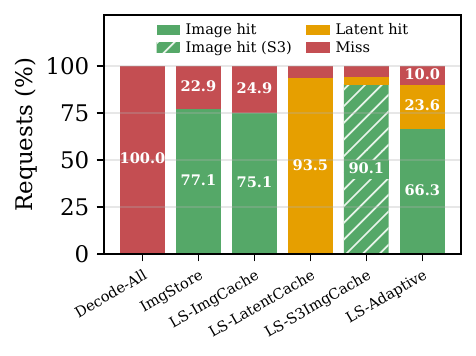}
\caption{\rev{Cache hit distribution.}}\label{fig:cache-hit-dist}
\end{subfigure}\hfill
\begin{subfigure}[t]{0.24\textwidth}
\centering
\includegraphics[width=\linewidth]{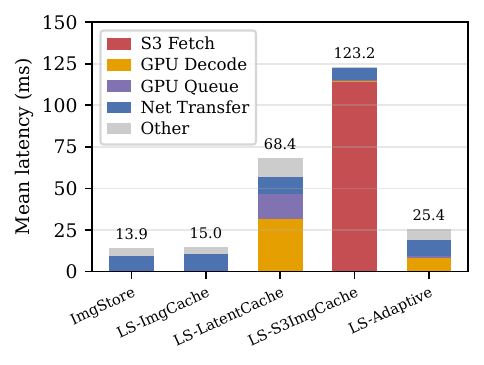}
\caption{\rev{Hit latency breakdown.}}\label{fig:latency-breakdown-hit}
\end{subfigure}\hfill
\begin{subfigure}[t]{0.24\textwidth}
\centering
\includegraphics[width=\linewidth]{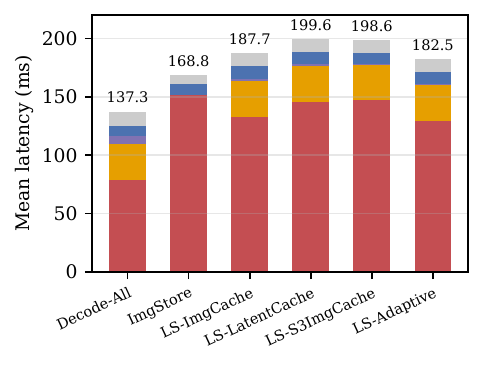}
\caption{\rev{Miss latency breakdown.}}\label{fig:latency-breakdown-miss}
\end{subfigure}
\vspace{-5pt}
\caption{End-to-end read performance.
  (a)~CDF of read latency for \rev{six} store-and-read configurations.
  (b)~Stacked image-hit, latent-hit, and full-miss fractions\rev{; hatched: image hits served from S3}.
  (c)~Mean latency breakdown for cache-hit requests (image hit + latent hit); numbers above bars show total mean (ms)\rev{; for LS-S3ImgCache, S3 Fetch is the GET of the cached PNG}.
  (d)~Same as (c) for cache-miss requests.}
\label{fig:e2e-performance}
\label{fig:latency-breakdown}
\vspace{-10pt}
\end{figure*}

\subsection{Storage Reduction}
\label{subsec:storage}
\rev{\textit{Storing compressed latents instead of PNGs should shrink the footprint severalfold, regardless of model or resolution}; \tref{tab:storage} tests this on 150\,K Stable Diffusion (SD)~3.5 and 100\,K FLUX.1 images at each of $512{\times}512$ and $1024{\times}1024$.}
\rev{\system attains an aggregate DRR of \textbf{78.7\%} even though ImgStore's PNGs are already losslessly compressed (against raw RGB, a $1024{\times}1024$ image shrinks from 3\,MB to ${\sim}$0.29\,MB).}
\rev{The encoder alone removes ${\sim}$65\% of ImgStore's bytes; pcodec then removes another 43\% (SD~3.5) or 34\% (FLUX.1) of the latent bytes, a gap likely due to their different VAEs.}
\rev{\textit{Latent-first storage thus removes over three quarters of the PNG bytes for both models at both resolutions}, so 100~billion $1024{\times}1024$ SD~3.5 images would take ${\sim}$30\,PB under \system instead of ${\sim}$140\,PB.}

\subsection{End-to-End Serving Performance}
\label{subsec:e2e}

We evaluate \system's read latency under the full 48-hour trace replay on the prototype cluster.
\rev{All store-and-read configurations (\sref{subsec:setup}) share the trace and 720-hour cache warmup.}

\subsubsection{Generation vs.\ Store-and-Read}
\rev{\textit{Regenerating each image on demand would avoid storing it, but should be too slow for serving}; \tref{tab:latency-summary} confirms this: generation averages over 4\,s per read, while Decode-All and \system are $31{\times}$ and $105{\times}$ faster.
\textit{Generated images or their latents must therefore be stored for fast access}; regeneration would also cost over an order of magnitude more (\sref{subsec:spectrum}).}

\begin{table}[t]
  \vspace{-5pt}
  \centering
  \caption{Read latency (ms). Store-and-read configurations use 3 nodes ${\times}$ 2\,GB per-node cache (6\,GB total, 66{,}581 requests). Generation latency is measured separately over 1,000 requests.}
  \label{tab:latency-summary}
  \vspace{-5pt}
  \scalebox{0.78}{\begin{tabular}{@{}lrrrr@{}}
  \toprule
  \textbf{Configuration} & \textbf{Mean} & \textbf{P50} & \textbf{P95} & \textbf{P99} \\
  \midrule
  Generation & 4{,}309 & 4{,}130 & 13{,}647 & 17{,}518 \\
  \midrule
  Decode-All       & 137.3 & 132.6 & 209.1 & 275.9 \\
  ImgStore         &  49.4 &  \textbf{14.1} & 198.7 & 276.9 \\
  LS-ImgCache      &  58.0 &  15.6 & 213.9 & 285.3 \\
  LS-LatentCache   &  77.0 &  53.8 & 180.9 & 234.3 \\
  \rev{LS-S3ImgCache} & \rev{127.9} & \rev{124.0} & \rev{234.0} & \rev{339.3} \\
  \textbf{LS-Adaptive} &  \textbf{41.2} &  15.8 & \textbf{178.5} & \textbf{227.3} \\
  \bottomrule
  \end{tabular}}
  \end{table}

\subsubsection{Overall Read Latency}
\fref{fig:latency-cdf} and \tref{tab:latency-summary} compare the \rev{six} store-and-read configurations. \rev{
ImgStore's image hits skip the GPU, so its median rivals \system's, but its 22.9\% misses fetch full PNGs from S3, inflating its mean and P99.
LS-ImgCache misses about as often but must also decode each miss; LS-LatentCache cuts misses to 6.5\% by caching ${\sim}5{\times}$ more objects, but decodes every hit, shifting its CDF rightward.
LS-S3ImgCache has no fast path: each of its ${\sim}$90\% hot-set hits is an S3 GET of a full PNG, so it is $3.6{\times}$ slower in mean than ImgStore with an equal-size DRAM cache and a similar hit rate (10\% in \tref{tab:imgstore-sweep}).
\system serves 66.3\% of reads as zero-decode image hits (\fref{fig:cache-hit-dist}) and most others as latent hits, so only 10.0\% go to S3.
\textit{\system thus keeps ImgStore's fast median yet has the lowest mean, P95, and P99 of the six}, with mean and P99 17\% and 18\% below ImgStore's.}

\subsubsection{Latency Breakdown and Cache Hit Distribution}
\label{subsubsec:breakdown}
\rev{\textit{The fastest design should avoid costly S3 misses without routing most reads through the GPU}; \fref{fig:latency-breakdown} tests this by splitting latency by request type.
A latent hit adds a GPU decode to an image hit (\fref{fig:latency-breakdown-hit}); a miss is dominated by its S3 fetch (\fref{fig:latency-breakdown-miss}) and costs ${\sim}3{\times}$ a latent hit and over $10{\times}$ an image hit.
A \system miss is no cheaper than an ImgStore miss, as the decode offsets its shorter fetch, but a ${\sim}5{\times}$ smaller latent fetch varies less than a full PNG fetch, lowering the tail.
LS-LatentCache decodes every request, so its latent hits queue ${\sim}4{\times}$ longer than \system's.
\textit{\system thus wins not through cheaper misses but by missing under half as often as ImgStore while keeping two thirds of reads off the GPU}.}

\subsubsection{\rev{Scaling Image Cache Capacity}}
\label{subsubsec:imgcache-scaling}

\begin{table}[t]
  \centering
  \caption{\rev{ImgStore with enlarged caches: image-hit rate and read latency (ms) vs.\ aggregate cache size (fraction of the 210.6\,GB unique-object footprint, summed over 3 nodes).}}
  \label{tab:imgstore-sweep}
  \vspace{-5pt}
  \scalebox{0.78}{\begin{tabular}{@{}lrrrrr@{}}
  \toprule
  \rev{\textbf{Cache}} & \rev{\textbf{Total}} & \rev{\textbf{Img-hit}} & \rev{\textbf{Mean}} & \rev{\textbf{P95}} & \rev{\textbf{P99}} \\
  \midrule
  \rev{ImgStore 1\%}  & \rev{2.1\,GB}  & \rev{62.0\%} & \rev{75.5} & \rev{228.5} & \rev{323.7} \\
  \rev{ImgStore 3\%}  & \rev{6.4\,GB}  & \rev{77.1\%} & \rev{49.4} & \rev{198.7} & \rev{276.9} \\
  \rev{ImgStore 5\%}  & \rev{10.7\,GB} & \rev{83.5\%} & \rev{45.5} & \rev{214.7} & \rev{296.8} \\
  \rev{ImgStore 10\%} & \rev{21.4\,GB} & \rev{90.6\%} & \rev{35.7} & \rev{196.2} & \rev{273.0} \\
  \rev{ImgStore 15\%} & \rev{32.1\,GB} & \rev{\textbf{93.8\%}} & \rev{\textbf{32.3}} & \rev{194.9} & \rev{269.7} \\
  \midrule
  \rev{\textbf{LS-Adaptive 3\%}} & \rev{6\,GB} & \rev{---} & \rev{41.2} & \rev{\textbf{178.5}} & \rev{\textbf{227.3}} \\
  \bottomrule
  \end{tabular}}
\end{table}

\rev{\textit{With a large enough image cache, ImgStore should match \system's latency without any decoding}; we sweep its cache from 1\% to 15\% of the footprint (\tref{tab:imgstore-sweep}).
The mean falls below \system's at roughly 7\% of the footprint, but the tail does not: residual misses still fetch full PNGs from S3, keeping P99 19--42\% above \system's at every size.
\textit{A larger image cache thus buys \system's mean latency but never its tail}; \sref{subsec:spectrum} prices that capacity.}
\subsection{Long-Term Cost Projection}
\label{subsec:cost-projection}

\rev{As stored images accumulate, \textit{\system's ${\sim}5{\times}$ smaller footprint should outweigh its decode and cache costs by a margin that grows with the horizon}; we test this by replaying the full 35-month \civitai trace through a cost model and extrapolating its steady state to 2050.}

\phead{Cost model.}
\rev{We count the cost components that differ across strategies: persistent storage, the cache, and on-demand GPU decoding. For costing, the dual-format cache runs on always-on nodes with local NVMe SSDs, billed per GB-month for the whole period, and GPU decoding is billed per decode~\cite{modal_pricing,runpod_serverless}; our testbed co-locates both on the GPU nodes.} Other common operational costs are omitted. \rev{Let $N(t)$ be the cumulative number of images at time $t$, $\bar{S}_{\text{px}}$ and $\bar{S}_{\text{lat}}$ the average PNG and compressed-latent sizes, $f$ the cache size as a fraction of the PNG working set, $M(t)$ the decode count, and $P_{\text{dec}} = t_{\text{dec}} \cdot P_{\text{GPU}}$ the per-decode cost.} Then
\vspace{-6pt}
\begin{align}
  C_{\text{ImgStore}}(t) &= N(t)\cdot \bar{S}_{\text{px}} \cdot P_{\text{S3}} \\
  C_{\text{\system}}(t) &= \underbrace{N(t)\cdot\bar{S}_{\text{lat}}\cdot P_{\text{S3}}}_{\text{latent storage}}
                   + \underbrace{f\cdot N(t)\cdot\bar{S}_{\text{px}}\cdot P_{\text{SSD}}}_{\text{SSD cache}} \nonumber \\
                   &\quad + \underbrace{M(t)\cdot P_{\text{dec}}}_{\text{GPU decode}}
  \label{eq:cost-model}
\vspace{-12pt} 
\end{align}
\rev{Our latency measurements (\sref{subsec:e2e}) use DRAM caches, whereas this model places caches on SSD, so the latencies we report for designs with SSD caches
are estimated from these measurements rather than measured. The cache medium matters only on cache hits. Reading a cached object from a local NVMe SSD instead of DRAM adds 3--5\,ms per hit. Mean latency therefore increases by at most that amount. Tail latency is set by misses, which bypass the cache and fetch from S3, and is essentially unaffected. We treat all designs with SSD caches alike, including \system itself.} 

\rev{\tref{tab:cost-params} lists all parameters. We compare four setups: ImgStore on S3 Standard; ImgStore + Glacier IR, which migrates objects older than 5 years to S3 Glacier Instant Retrieval~\cite{glacier_ir} and models retrieval cost with the stratified age-decay fit from O2; and LS-Adaptive with a 1\% dual-format cache on H100 or RTX~5090 GPUs, where $m_{\text{gpu}}{=}63.2\%$ of requests trigger a decode in the last 12 months of a full-trace simulation. Per-decode billing lower-bounds GPU cost: with an always-on GPU fleet sized for the peak minute instead, the H100 saving vanishes, while RTX~5090 still saves 44\% by 2050.}

\begin{table}[t]
\centering
\caption{Cost model parameters.}
\label{tab:cost-params}
\vspace{-7pt}
\resizebox{\linewidth}{!}{\begin{tabular}{@{}lrl@{}}
\toprule
\textbf{Parameter} & \textbf{Value} & \textbf{Notes} \\
\midrule
$\bar{S}_{\text{px}}$  & 1.5\,MB    & Average PNG ($1024 \times 1024$) \\
$\bar{S}_{\text{lat}}$ & 0.29\,MB   & Compressed latent (SD\,3.5) \\
\midrule
$P_{\text{S3}}$        & \$0.023/GB-mo & AWS S3 Standard~\cite{aws_s3_pricing} \\
$P_{\text{Glacier}}$   & \$0.004/GB-mo & S3 Glacier IR storage (objects $\geq$5\,yr)~\cite{glacier_ir} \\
$P_{\text{GIR-ret}}$  & \$0.01/GB + \$0.0001/req & S3 Glacier IR retrieval \\
$P_{\text{H100}}$      & \$2.50/GPU-hr & Datacenter GPU rental~\cite{jarvislabs2026h100price} \\
$P_{\text{5090}}$      & \$0.69/GPU-hr & Consumer-class GPU rental~\cite{5090pricing} \\
$t_{\text{dec}}$       & 40\,ms        & SD\,3.5 decode \\
\rev{$P_{\text{DRAM}}$} & \rev{\$2.49/GB-mo} & \rev{Marginal EC2 DRAM (r6i$-$c6i)~\cite{aws_ec2_pricing}} \\
\rev{$P_{\text{SSD}}$}  & \rev{\$0.132/GB-mo} & \rev{Local NVMe SSD cache node (i3en)~\cite{aws_ec2_pricing}} \\
\midrule
$f$                    & 1\%        & \rev{Cache size as fraction of working set} \\
$m_{\text{gpu}}$       & 63.2\%     & \system{} decode-trigger rate \\
$\lambda$              & 10.2/yr    & Mean views per image \\
\bottomrule
\end{tabular}}
\end{table}

\begin{figure}[t]
  \centering
  \scalebox{0.75}{\includegraphics[width=\columnwidth]{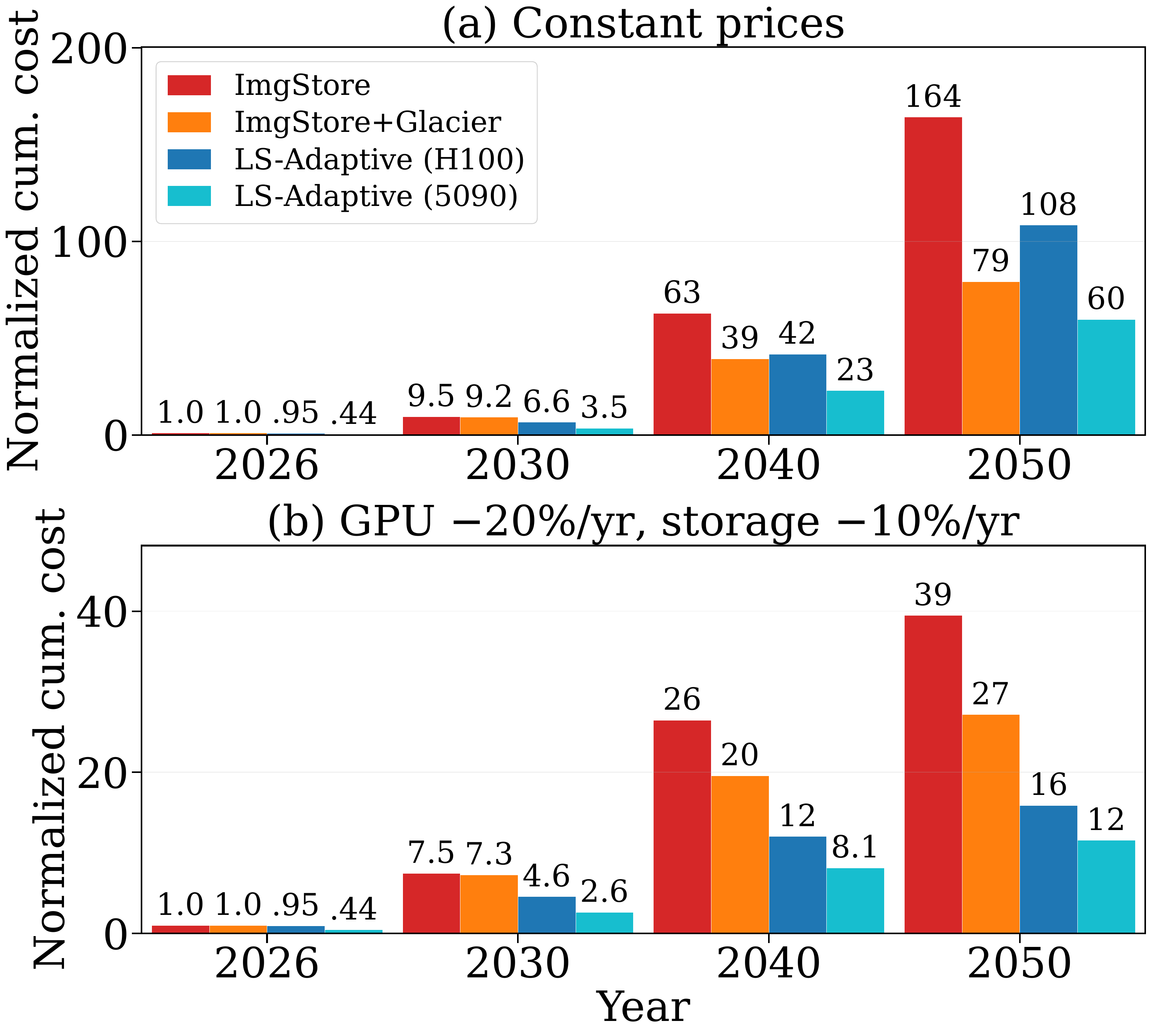}}
  \vspace{-8pt}
  \caption{Cumulative cost at four time horizons (2026, 2030, 2040, 2050), normalized so ImgStore at the trace end (March~2026) equals~1. (Top)~constant prices. (Bottom)~with annual price decay (GPU $-$20\%/yr, storage $-$10\%/yr from 2026~\cite{mccallum2023diskprice,diskprices2025,epochai2024mlhardware,epochai2024priceperf}).}
  \label{fig:cumulative-cost}
\end{figure}

\rev{The projection extrapolates the trace tail's steady state ($\sim$3.76\,M new images per month, a 12.7\% CAGR of the cumulative working set over the horizon) with $m_{\text{gpu}}$ and $\lambda$ held constant, so monthly decode demand is $m_{\text{gpu}}\cdot\lambda\cdot N(t)/12$; \fref{fig:cumulative-cost} plots the resulting cumulative cost.}

\phead{Cost during the trace period.}
\rev{Over the 35-month trace, LS-Adaptive already saves \textbf{56\%} on 5090 but only 5\% on H100, whose higher GPU rate, together with the cache, offsets most of the still-modest storage gap on this short horizon.}

\phead{Long-term projection.}
\rev{By 2050, storage has accumulated over the linearly growing working set and \system's footprint advantage becomes decisive: at constant prices (\fref{fig:cumulative-cost}-top), LS-Adaptive saves \textbf{64\%} on 5090 and 34\% on H100. Glacier IR tiering roughly halves ImgStore's cost, yet LS-Adaptive on 5090 still saves 25\% against it, whereas on H100 it loses to the tiered baseline at constant prices.}

\rev{Under the optimistic price-decline scenario (\fref{fig:cumulative-cost}-bottom), the gap widens: by 2050 LS-Adaptive on 5090 saves \textbf{71\%} over ImgStore and 57\% over ImgStore + Glacier IR, and even H100 beats the tiered baseline by 42\%, because cheaper GPUs amplify the decode-based design's advantage while cheaper storage benefits both proportionally. \textit{\system's saving thus grows with the horizon and with cheaper GPUs, and on RTX~5090 it beats even a Glacier-tiered ImgStore}.}

\subsection{\rev{Storage--Compute Tradeoff}}
\label{subsec:spectrum}

\begin{figure}[t]
  \centering
  \includegraphics[width=0.72\columnwidth]{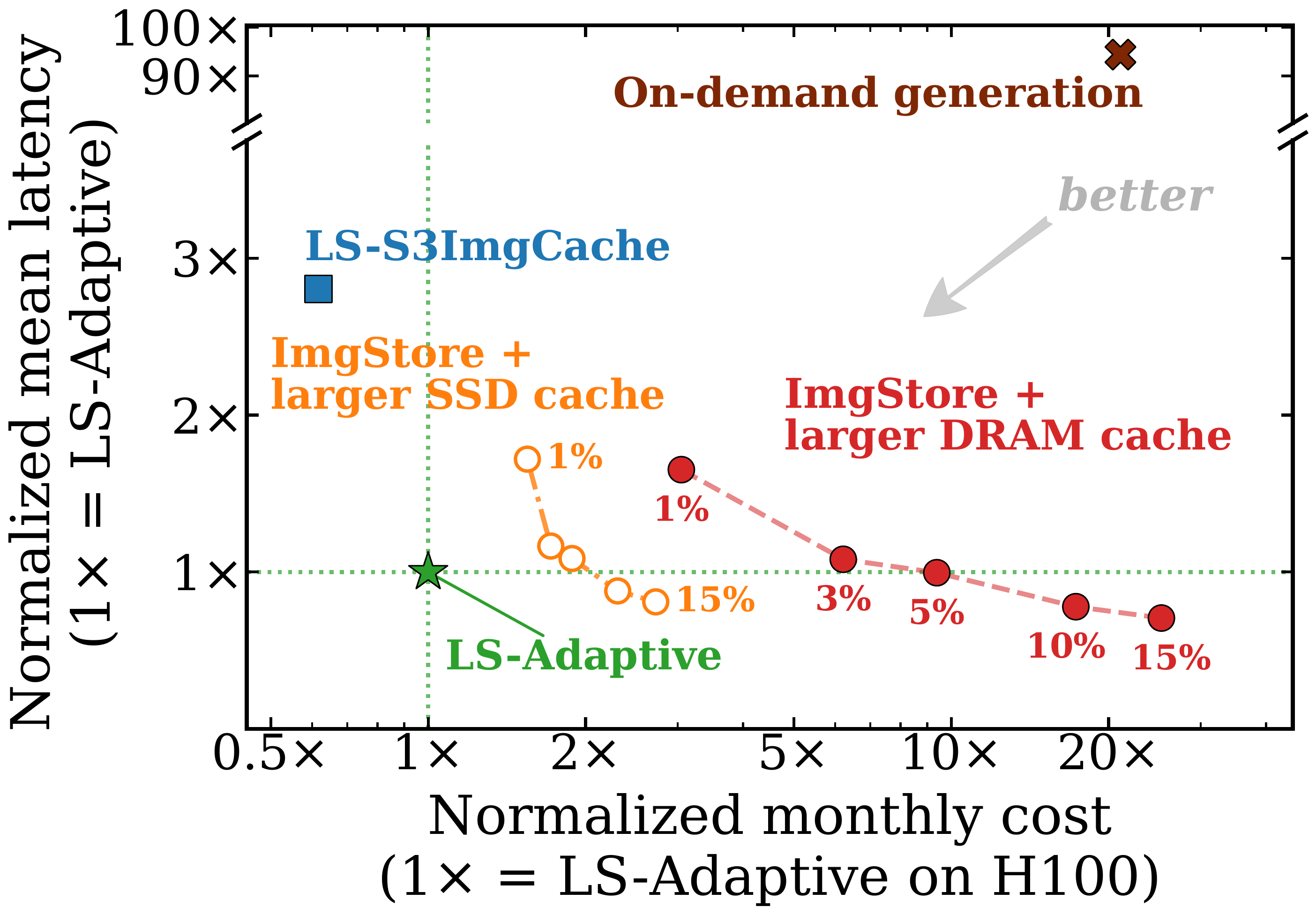}
  \vspace{-10pt}
  \caption{\rev{Monthly cost vs.\ mean read latency of the storage--compute alternatives, normalized to LS-Adaptive (H100 pricing). Latencies of SSD-cached designs, including LS-Adaptive, are estimated from DRAM measurements (\sref{subsec:cost-projection}).}}
  \label{fig:cost-latency-plane}
\end{figure}

\rev{\textit{Every alternative that shifts the storage--compute balance away from \system's should lose on cost, latency, or both}; \fref{fig:cost-latency-plane} tests this by placing them all on one cost--latency plane, with costs from the model of \sref{subsec:cost-projection} at the end-of-trace platform scale and mean latencies from \sref{subsec:e2e}, estimated for SSD caches.}

\rev{Of these alternatives, only LS-S3ImgCache is cheaper than LS-Adaptive, by 38\% under H100 pricing, yet every hit is an S3 round trip, so it is nearly $3{\times}$ slower on the mean and $1.5{\times}$ on P99. With every design's cache charged per GB-month here (\tref{tab:cost-params}), ImgStore with an enlarged image cache fails on cost instead: a cache large enough to match LS-Adaptive's mean (\sref{subsubsec:imgcache-scaling}) costs $9{\times}$ LS-Adaptive's total in DRAM and still $2{\times}$ on SSD, because ImgStore's S3 PNG storage alone exceeds LS-Adaptive's total. On-demand generation costs more than an order of magnitude more and is two orders of magnitude slower than LS-Adaptive. \textit{No alternative therefore matches \system on both axes: the only cheaper one is nearly $3{\times}$ slower, and those that match its mean cost about twice as much or more and still have a higher tail.}}

\subsection{Ablation and Sensitivity Analysis}
\label{subsec:ablation}

We use both prototype-cluster experiments and trace-driven simulation to validate each of \system's three core design choices and to characterize parameter sensitivity:
\begin{enumerate}[noitemsep,leftmargin=*]
  \item \textbf{Dual-format caching} vs.\ single-format baselines (\S\ref{subsubsec:cache-size});
  \item \textbf{Online ratio tuning} vs.\ static allocation (\S\ref{subsubsec:static-comparison});
  \item \textbf{Spillover dispatch} vs.\ hash-only routing (\S\ref{subsubsec:spillover});
  \item \textbf{Parameter sensitivity} of the tuning algorithm (\S\ref{subsubsec:param-sensitivity}).
\end{enumerate}
For the simulation experiments, the simulator faithfully implements the dual-format cache with independent LRU tiers, the main/tail split, and the marginal-hit gradient update rule.
The tuning algorithm uses six parameters: $T_{decode}$ and $T_{fetch}$ are the per-request latency costs of a latent-tier hit (GPU decode) and a full miss (S3 fetch + decode), respectively; $\Delta$ is the step size by which $\alpha$ is adjusted at each window boundary; $W$ is the number of requests per gradient-estimation window; $\tau$ is the fraction of each LRU tier reserved as a \emph{tail} segment for marginal hit-rate estimation; and $h$ is the promotion threshold, i.e., the number of latent hits an object must accumulate before being promoted to the image tier. Unless otherwise noted, simulations use $T_{decode}{=}40$\,ms, $T_{fetch}{=}140$\,ms, $\Delta{=}0.005$, $W{=}1{,}000{,}000$, $\tau{=}0.10$, and $h{=}8$.

\subsubsection{Dual-Format Cache across Cache Sizes}
\label{subsubsec:cache-size}

\begin{table}[t]
  \centering
  \caption{Latency (ms) across cache sizes (\% of WSS) for \system's dual-format cache vs.\ single-format baselines.}
  \label{tab:cache-size-comparison}
  \vspace{-5pt}
  \scalebox{0.7}{\begin{tabular}{@{}lrrrrrr@{}}
  \toprule
   & \multicolumn{6}{c}{\textbf{Cache size (\% of WSS)}} \\
   \cmidrule(l){2-7}
  \textbf{Configuration} & \textbf{0.1\%} & \textbf{0.5\%} & \textbf{1\%} & \textbf{2\%} & \textbf{5\%} & \textbf{10\%} \\
  \midrule
  LS-ImgCache      & 103.1 & 73.9 & 61.0 & 48.7 & 33.7 & 23.2 \\
  LS-LatentCache   &  97.5 & 74.9 & 66.2 & 58.0 & 50.1 & 47.0 \\
  \textbf{LS-Adaptive} & \textbf{90.1} & \textbf{62.2} & \textbf{50.7} & \textbf{40.1} & \textbf{28.5} & \textbf{21.9} \\
  \bottomrule
  \end{tabular}}
\end{table}

\tref{tab:cache-size-comparison} compares \system's dual-format cache against its two single-format variants, LS-ImgCache ($\alpha{=}1$) and LS-LatentCache ($\alpha{=}0$), across six cache sizes.
At the smallest cache size (0.1\%), LS-LatentCache is competitive because compressed latents are ${\sim}5{\times}$ smaller than PNGs, fitting more objects into the same budget.
As the cache grows, repeated GPU decode on popular items becomes the bottleneck; LS-ImgCache surpasses LS-LatentCache starting at 0.5\%.
The best single-format policy thus changes with cache size, confirming that no single format dominates.
\system's dual-format cache achieves the \textbf{lowest} latency at every capacity, outperforming the best single-format alternative by 5--18\%.

\subsubsection{Online Tuning vs.\ Static Allocation}
\label{subsubsec:static-comparison}

\begin{figure}[t]
\centering
\begin{minipage}[t]{0.54\columnwidth}
\centering
\includegraphics[width=\linewidth, height=4.5cm, keepaspectratio]{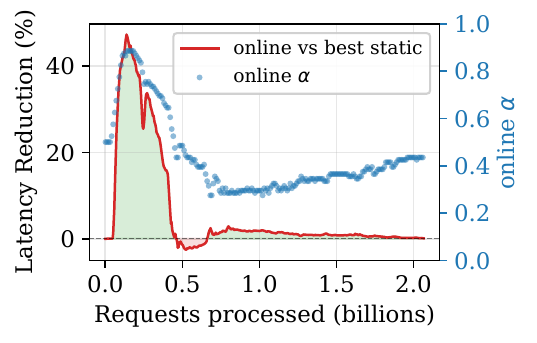}
\vspace{-20pt} 
\captionof{figure}{~Per-window latency improvement and $\alpha$ trajectory.} 
\label{fig:static-vs-online}
\end{minipage}\hfill
\begin{minipage}[t]{0.42\columnwidth}
\centering
\includegraphics[width=\linewidth, height=4.5cm, keepaspectratio]{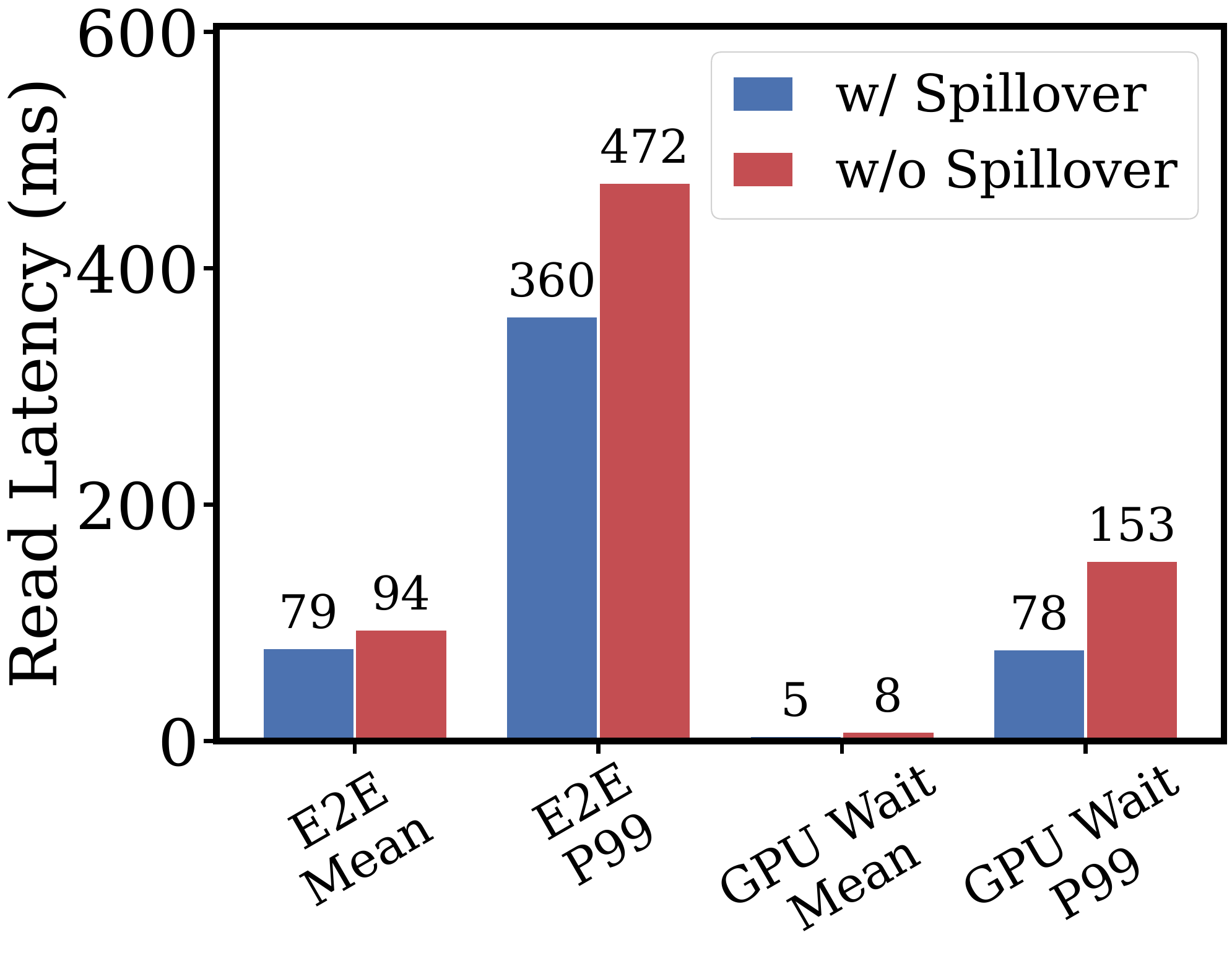}
\vspace{-20pt}
\captionof{figure}{Latency with and without spillover dispatch.}
\label{fig:spillover-ablation}
\end{minipage}
\vspace{-3pt}
\end{figure}

A natural question is whether a \emph{fixed} $\alpha$, picked offline, could match the adaptive resizer.
We sweep $\alpha \in \{0.3, 0.4, 0.5, 0.6, 0.7\}$ on the full 2.07\,B-request trace; the best static choice is $\alpha{=}0.5$ at 51.6\,ms. \system's adaptive resizer reaches \textbf{50.7\,ms}, 1.7\% lower than this oracle-picked best static, and matches or beats $\alpha{=}0.5$ in \textbf{87\%} of all 1\,M-request windows, without prior knowledge of the trace.

\fref{fig:static-vs-online} shows where the gain comes from. During the first 0.4\,B requests, the 1\% budget already covers most of the small catalog of up to 4\,M images, so format choice dominates over capacity. \system raises $\alpha$ to 0.7--0.85 to prioritize image hits that skip the 40\,ms decode, while any fixed $\alpha{\leq}0.5$ wastes half its slots on latents that still pay it; the gain peaks at \textbf{30--45\%}.
Once the catalog grows past 60\,M, misses dominate latency regardless of $\alpha$ and the gain settles into a steady-state \textbf{1--3\%} lead, with $\alpha$ pulled back to 0.3--0.4. The single brief dip below zero near 0.5\,B requests is the phase transition where the catalog first outgrows the cache and the controller takes a few windows to re-converge.

The practical message is twofold. First, even on this stationary trace where a single static $\alpha$ is competitive, the adaptive resizer is strictly better on average and never asks the operator to guess. Second, the cost of guessing wrong is real: $\alpha{=}0.3$ and $\alpha{=}0.7$ are both within 2.4\% of the best static on this trace, but would drift from the optimum on a workload whose hit-cost distribution is even modestly different, e.g., after a model migration that changes $T_{decode}$ or a content surge that shrinks cache coverage. \system absorbs that risk automatically.

\subsubsection{Spillover Dispatch}
\label{subsubsec:spillover}

To validate the effectiveness of the spillover path, 
we replay the same 48-hour trace on a 6-node GPU cluster at $1000{\times}$ speed and an overflow threshold $\theta{=}4$.
The \emph{without-spillover} baseline sets $\theta$ to infinity, so every request is dispatched to its hash-determined owner node regardless of queue depth.
\fref{fig:spillover-ablation} compares latency with and without spillover.
Spillover reduces mean E2E latency by \textbf{16.5\%} (94.5\,ms to 78.9\,ms) and P99 by \textbf{23.9\%} (472.5\,ms to 359.5\,ms).

The gain is concentrated in GPU queue wait time, where P99 drops by 49\% from 152.8\,ms to 78.0\,ms, confirming that spillover alleviates head-of-line blocking on hot nodes.
Under consistent-hashing routing, natural load skew causes some nodes to accumulate deeper queues; spillover detects this via the threshold $\theta$ and redirects excess work to idle nodes while keeping cache entries pinned to their owner nodes, preserving locality without sacrificing tail latency.
\rev{Spillover is rare and cheap: only ${\sim}$3\% of requests spill, each moving at most one compressed latent (${\sim}$0.3\,MB) and one write-back (${\sim}$1.4\,MB); even at $1000{\times}$ speed, the peak spill traffic (spill count times payload) is ${\sim}$50\,MB/s, far below the 25\,Gbps inter-node link.}

\begin{figure}[t]
  \centering
  \includegraphics[width=\columnwidth]{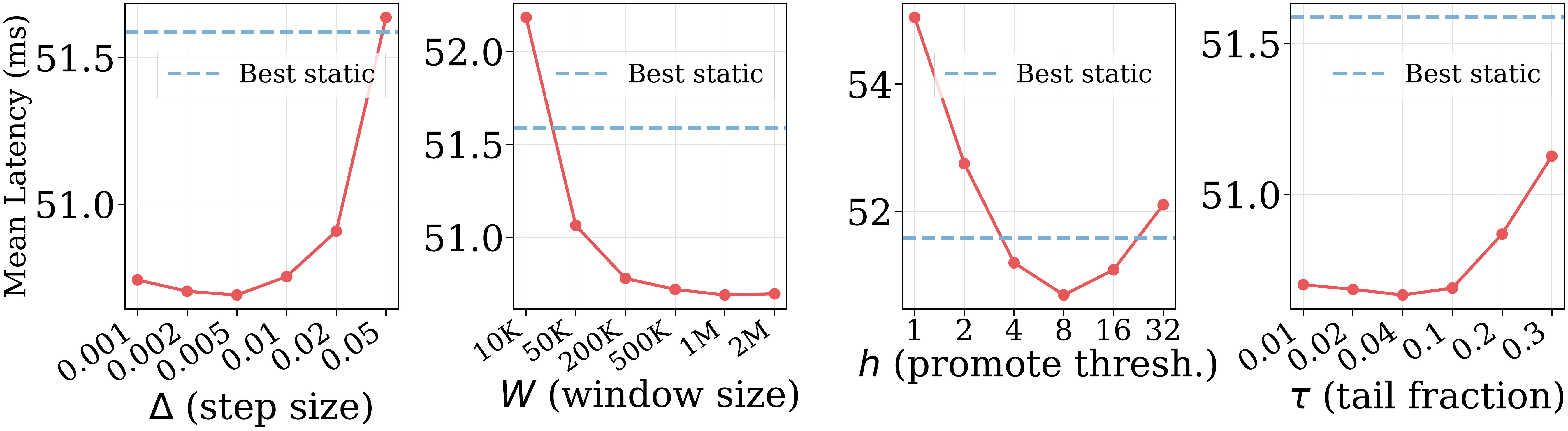}
  \caption{Sensitivity analysis. \system is robust to step size, window size, and tail fraction; promotion threshold $h$ matters most.}
  \label{fig:param-sensitivity}
\end{figure}

\subsubsection{Parameter Sensitivity}
\label{subsubsec:param-sensitivity}

\fref{fig:param-sensitivity} examines the sensitivity of the online tuning to its four main parameters: step size $\Delta$, window size $W$, promotion threshold $h$, and tail fraction $\tau$.
We sweep each parameter across a range while holding the others at their defaults ($\Delta{=}0.005$, $W{=}1$M, $\tau{=}0.10$, $h{=}8$), and report the latency on the full \civitai trace at 1\% cache size.

The results show that \system is robust to parameter choices across three of the four parameters.
\textbf{Step size $\Delta$} (0.001--0.05): latency varies within 1.9\% of the best; values from 0.001 to 0.02 are nearly identical ($<$0.5\% spread), with only $\Delta{=}0.05$ showing a noticeable penalty.
\textbf{Window size $W$} (10K--2M): large windows ($\geq$200K) achieve near-optimal latency (within 0.2\%); smaller windows ($W{=}10$K) incur a 3\% penalty due to noisier gradient estimates.
\textbf{Tail fraction $\tau$} (0.01--0.3): the most insensitive parameter, with the entire range spanning only 0.9\%.
\textbf{Promotion threshold $h$} has the largest effect (8.6\% spread). Small values ($h{=}1$) promote aggressively, consuming image-cache capacity on rarely reused items; $h{=}8$ is optimal, and $h{\geq}4$ brings latency within 1\% of the minimum. Overly large values ($h{=}32$) delay promotion excessively, increasing latent-tier decode cost.

The insensitivity of $\Delta$, $W$, and $\tau$ follows from the gradient's self-correcting nature: an overly large step overshoots the optimal $\alpha$ but is corrected in the next window. \system can thus be deployed without extensive parameter tuning.

\subsection{Comparing with Lossy Compression}
\label{subsec:fidelity}

\begin{figure}[t]
    \centering
    \begin{subfigure}[t]{0.33\columnwidth}
    \includegraphics[width=\linewidth]{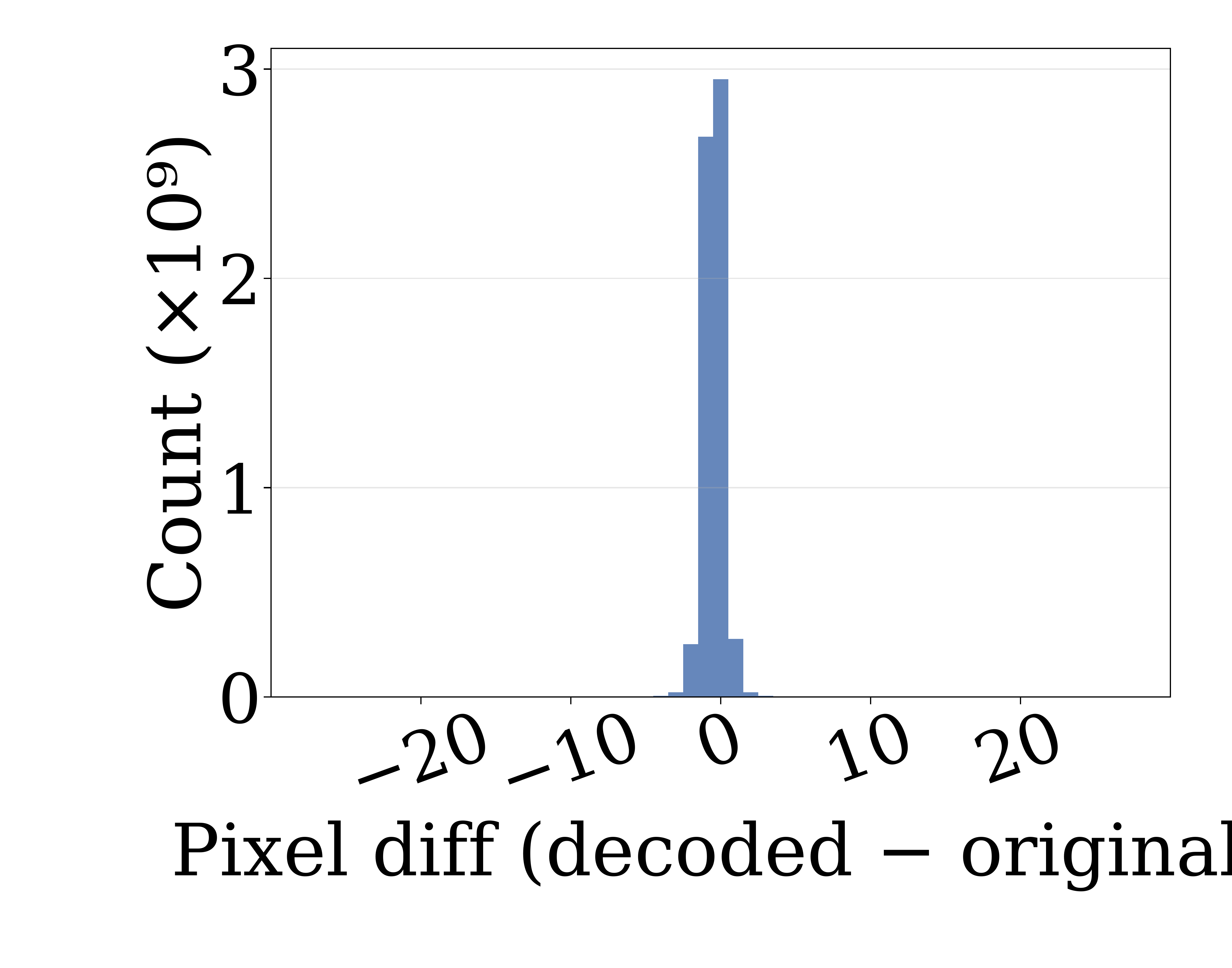}
    \caption{Pixel diff distribution.}\label{fig:pixel-diff}
    \end{subfigure}\hfil
    \begin{subfigure}[t]{0.33\columnwidth}
    \includegraphics[width=\linewidth]{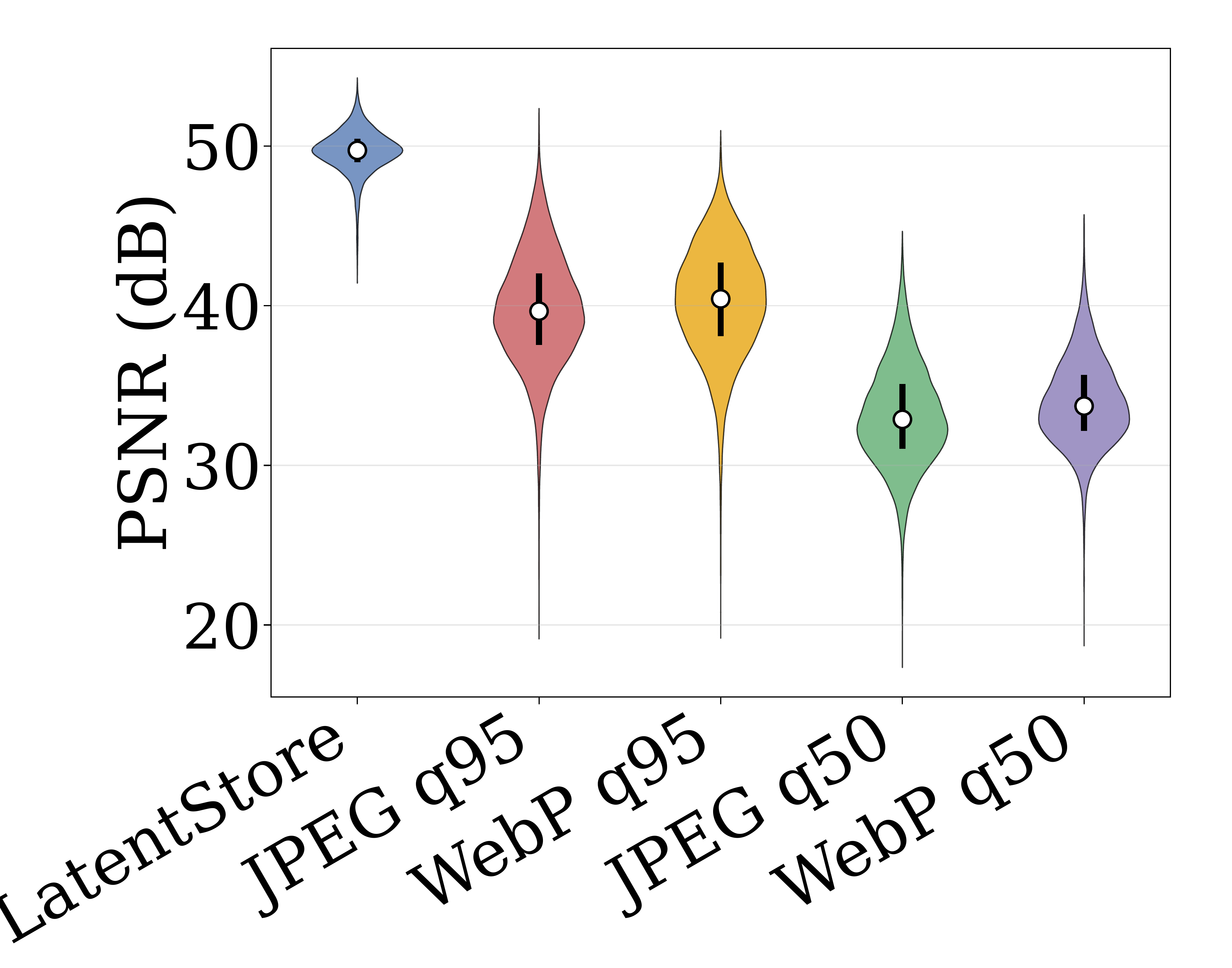}
    \caption{PSNR distribution.}\label{fig:violin-psnr}
    \end{subfigure}\hfil
    \begin{subfigure}[t]{0.33\columnwidth}
    \includegraphics[width=\linewidth]{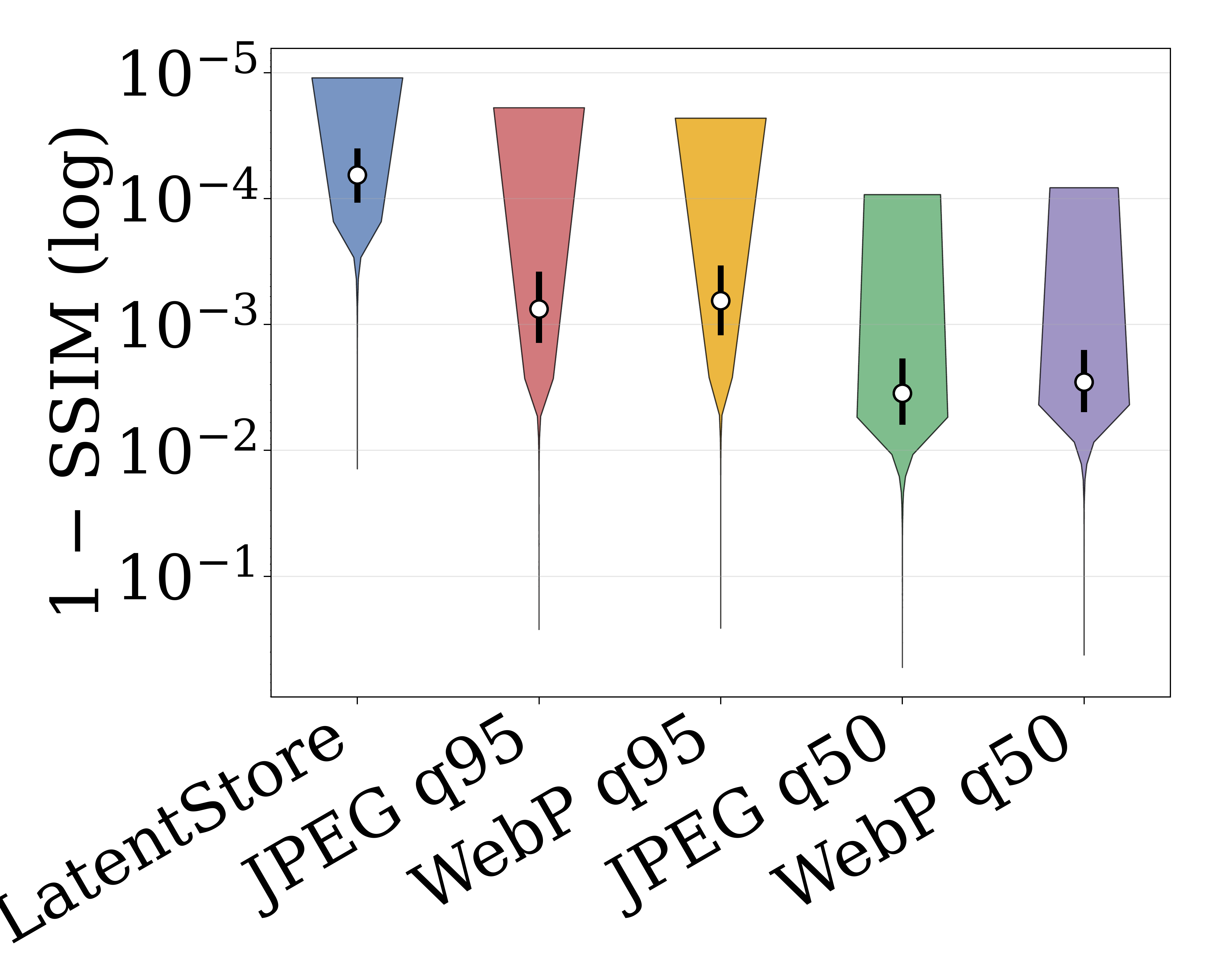}
    \caption{SSIM distribution.}\label{fig:violin-ssim}
    \end{subfigure}
    \caption{Reconstruction fidelity over 10\,K SD~3.5 images ($1024{\times}1024$). (a)~Signed per-channel pixel difference over 2{,}000 sampled images; 47\% of pixel-channel values are unchanged. (b)-(c)~White dot: median; thick bar: IQR. ``q'' in ``q95'' denotes quality factor; higher PSNR and SSIM indicate better fidelity.}  
    \label{fig:quality-violin}
    \end{figure}

\phead{Decode determinism.}
\rev{\textit{\system's savings should not come from discarding information: a latent should reproduce its pixels exactly under a pinned decoder and, even on other hardware, stay more faithful than a same-size lossy codec.}
The first holds by construction: latents are stored bit-exactly (pcodec is lossless) and VAE decoding is deterministic.
Across hardware, floating-point rounding differs: between H100 and L4, 47\% of pixel-channel values stay bit-exact and most others shift by $\pm$1--3 of 255 levels, symmetrically around zero (\fref{fig:pixel-diff}), i.e., unbiased rounding noise.
\system therefore tags each latent with its (model ID, decoder version, precision profile) and archives the decoder weights (${<}100$\,MB per family), so a latent is always decoded by the configuration that produced it; reads leaving the platform are served as pixels.}

\phead{Comparison with lossy codecs.}
\rev{We compare this noise with the error of lossy codecs using PSNR~\cite{gonzalez2018digital} and SSIM~\cite{hore2010image}; \system's 290\,KB latent falls between WebP\,q95 (260\,KB) and JPEG\,q95 (333\,KB).
\system's mean PSNR is 49.7\,dB versus 40.3\,dB for WebP\,q95, the best lossy codec, with SSIM in agreement (\fref{fig:violin-psnr}, \fref{fig:violin-ssim}); unlike the lossy codecs, \system has no long lower tail on textured or sharp-edged images.
\textit{A latent is thus exact under its pinned decoder and, even decoded elsewhere, ${\sim}$10\,dB more faithful (${\sim}10{\times}$ lower MSE) than a q95 JPEG or WebP file of its size: \system's savings do not come from discarding information.}}

\section{Related Work}
\label{sec:related}

\system lies at the intersection of large-scale image storage, generative-model optimization, computation--storage tradeoffs, and cache management.
We discuss how it relates to and departs from prior work in each area.

\phead{Large-scale image storage systems.} 
Conventional object storage systems~\cite{beaver2010finding, muralidhar2014f4, noghabi2016ambry, tencent_trace, calder2011windows, balakrishnan2014pelican} have been extensively studied for serving billions of blobs. However, these systems treat images as opaque pixel blobs, whereas \system stores compact latents and reconstructs pixels on demand.

\phead{Diffusion model inference optimizations.} 
Extensive research accelerates image generation via efficient sampling~\cite{song2021ddim, liu2022pndm}, optimized architectures~\cite{chen2023pixart, karras2022edm}, and inference-time scaling~\cite{ma2025inference}. Recent work further reduces latency through few-step distillation~\cite{luo2023latent, yin2024one}, feature caching~\cite{ma2024deepcache}, and quantization~\cite{zheng2024bidm}. However, these generation-stage optimizations---mapping noise to latents---are orthogonal to \system. 
Faster generation does not ease the compounding storage burden of retaining billions of synthetic images. \system operates strictly \emph{after} generation, focusing on efficiently storing, caching, and reconstructing the resulting latents for serving.

\phead{Cache management and adaptive partitioning.}
A line of work optimizes co-managing compressed and uncompressed data in the same memory cache~\cite{wilson1999compressed, tuduce2005adaptive, zexpander_eurosys16}. 
Related work also explores storage reduction for AI models through compression and deduplication~\cite{su2024everything,wang2026zipllm,lan2026tensordex}.
Google's software-defined far memory~\cite{lagarcavilla2019farmemory} and Meta's TMO~\cite{weiner2022tmo} demote cold pages to a compressed tier in datacenters and pair a hot uncompressed tier with a cold compressed tier sized by page-pressure signals, trading CPU decompression for a larger effective footprint.   
Unlike these systems, \system does not tier the same data representation across compressed and uncompressed forms, but instead manages two distinct storage formats---decoded images and compressed latents. 

\par
\begingroup
\titlespacing{\section}{0pt}{6pt plus 1pt minus 1pt}{4pt}
\section{Conclusion}
\label{sec:conclusion}

This paper presents {\system}, a novel latent-first storage system. 
\system stores images as compressed latents and reconstructs pixels on demand using sparse GPU decoding, trading inexpensive compute for large persistent storage savings. To keep decode overhead low, \system uses a small dual-format cache that balances an image LRU tier for fast hits against a latent LRU tier for more effective capacity. Evaluated with a large-scale production trace, \system substantially reduces persistent storage cost while also lowering mean and tail read latency.
More broadly, this work suggests a shift in storing generated content. As generative platforms scale and GPU costs fall, the compute-for-storage tradeoff offers a practical path toward sustainable media storage. 

\par
\endgroup

\clearpage \section*{Acknowledgment}
\label{sec:acknowledgment } 
We thank our shepherd, Nitin Agrawal, and the anonymous reviewers for their valuable feedback and comments, which improved the paper. We thank Justin Maier, CEO of Civitai, and the Civitai team for insightful discussions and feedback on this work. This research was supported in part by National Science Foundation (NSF) grants CNS-2322860, OAC-2411009, and OAC-2403313. We also acknowledge support from NSF CloudBank for providing AWS credits and the Commonwealth Cyber Initiative (CCI).

\bibliographystyle{plain}
\bibliography{reference}

\end{document}